\documentclass[letterpaper,12pt]{article}
\pdfoutput=1
\usepackage{jheppub}
\usepackage{slashed}
\usepackage{graphicx} 
\usepackage{amsmath} 
\usepackage{multirow} 
\usepackage{hyperref}
\usepackage{epstopdf}
\usepackage{mathtools}
\usepackage{romannum}
\usepackage{color}
\usepackage{tikz-feynman}
\usepackage{float}

\usepackage{url}
\usepackage{slashed,stmaryrd}
\usepackage{caption}
\usepackage{subcaption}
\usepackage{xcolor}
\usepackage{makecell}
\usepackage[utf8]{inputenc}

\newcommand{\beq}{\begin{equation}}
\newcommand{\eeq}{\end{equation}} 
\newcommand{\bea}{\begin{eqnarray}}
\newcommand{\eea}{\end{eqnarray}}
\newcommand{\ba}{\begin{array}} 
\newcommand{\ea}{\end{array}}

\RequirePackage[normalem]{ulem} 
\RequirePackage{color}\definecolor{RED}{rgb}{1,0,0}\definecolor{BLUE}{rgb}{0,0,1} 

\def\m1{M_1}
\def\m2{M_2}
\def\m3{M_3}

\def\ch10{\tilde \chi^0_1}

\def\to{\rightarrow}

\newcommand{\lsim}{\mathrel{\mathop{\kern 0pt \rlap
  {\raise.2ex\hbox{$<$}}}
  \lower.9ex\hbox{\kern-.190em $\sim$}}} 
\newcommand{\gsim}{\mathrel{\mathop{\kern 0pt \rlap
  {\raise.2ex\hbox{$>$}}}
  \lower.9ex\hbox{\kern-.190em $\sim$}}}

\definecolor{pink}{RGB}{255,105,180}

\definecolor{ywgreen}{rgb}{0.09, 0.45, 0.27}

\title{Electroweak $\!$phase $\!$transition $\!$with $\!$spontaneous $\!$\boldmath{$Z_2$}-breaking}

\author[a,b,c]{Marcela Carena,}
\author[d]{Zhen Liu}
\author[a,b]{and Yikun Wang}

\affiliation[a]{Theoretical Physics Department, Fermi National Accelerator Laboratory, Batavia, Illinois, 60510, USA}
\affiliation[b]{Enrico Fermi Institute, University of Chicago, Chicago, Illinois, 60637, USA}
\affiliation[c]{Kavli Institute for Cosmological Physics, University of Chicago, Chicago, Illinois, 60637, USA}
\affiliation[d]{Maryland Center for Fundamental Physics, Department of Physics,
University of Maryland, College Park, MD 20742, USA}

\emailAdd{carena@fnal.gov}
\emailAdd{zliuphys@umd.edu}
\emailAdd{yikwang@uchicago.edu}

\abstract{

This work investigates a simple, representative extension of the Standard Model with a real scalar singlet and spontaneous $Z_2$ breaking, which allows for a strongly first-order phase transition, as required by electroweak baryogenesis. We perform analytical and numerical calculations that systematically include one-loop thermal effects, Coleman-Weinberg corrections, and daisy resummation, as well as evaluation of bubble nucleation. We study the rich thermal history and identify the conditions for a strongly first-order electroweak phase transition  with nearly degenerate extrema at zero temperature. This requires a light scalar with mass below 50 GeV.  Exotic Higgs decays, as well as 
  Higgs coupling precision measurements  at the LHC and future collider facilities, will test this model. Additional information may be obtained from future collider constraints on the Higgs self-coupling. 
Gravitational-wave signals are typically too low to be probed by future gravitational wave experiments.
}


\preprint{
\begin{flushright}
FERMILAB-PUB-19-602-T
\end{flushright}
}

\begin{document}
\maketitle
\flushbottom 

\section{Introduction}
\label{sec:intro}

%


The observed electroweak phase transition (EWPhT) in nature, together with sufficient  C and CP violation, provides one of the most appealing opportunity to solve the puzzle of the matter-antimatter asymmetry of the universe, namely the mechanism of Electroweak Baryogenesis (EWBG)~\cite{Klinkhamer:1984di,Kuzmin:1985mm,Arnold:1987mh,Arnold:1987zg,Khlebnikov:1988sr,Kajantie:1996mn,Riotto:1999yt}. 
For a successful EWBG, the EWPhT needs to be strongly first-order to create an out-of-equilibrium condition, and to assure that the baryonic asymmetry generated during the bubble nucleation is not erased by the sphaleron processes~\cite{Sakharov:1967dj, Kuzmin:1985mm}. In the Standard Model (SM) of particle physics, however, the EWPhT is a smooth crossover~\cite{Kajantie:1996mn, Kajantie:1996qd}, and the amount of CP violation is insufficient~\cite{Kuzmin:1985mm}, hence precluding the possibility of baryogenesis at the electroweak scale.  Extensions to the SM that enhance the EWPhT and provide new sources of CP violation offer new possibilities for EWBG and have been studied extensively in the literature, e.g.~see~\cite{Chung:2012vg,Morrissey:2012db}.

Singlet extensions of the SM provide a unique opportunity to generate a  strongly first-order EWPhT \cite{Espinosa:1993bs,Espinosa:2007qk,Profumo:2007wc,Espinosa:2011ax,Barger:2011vm,Curtin:2014jma,Kurup:2017dzf,Choi:1993cv,Ham:2004cf,Noble:2007kk,Ashoorioon:2009nf,Das:2009ue,Profumo:2014opa,Kotwal:2016tex,Benson:1993qx,McDonald:1993ey,Ahriche:2007jp,Espinosa:2008kw,Chen:2017qcz,Barger:2008jx,Tenkanen:2016idg,Ghorbani:2018yfr}, and are the subject of exploration in this work. These extensions, however, are relatively difficult to test, in comparison with other SM particle extensions  with particles charged under the SM gauge groups. 
On the other hand, dark sector model building, involving a hidden sector with dark matter, often invokes spontaneously broken dark gauge symmetries.
The simplest scalar sector charged under the dark symmetries would be a complex scalar, which is a singlet under the SM gauge groups. The effects from the dark Higgs, which obtains a  vacuum expectation value (VEV), on the EWPhT can be approximated by a singlet extension of the SM with spontaneous $Z_2$ breaking, after rescaling the parameters by the corresponding degrees of freedoms.
 Given the above picture, in this work, we consider a  comparative study of a real singlet extension of the SM and its impact on the strength of the EWPhT,  in the presence of spontaneous Z2 breaking, through a  detailed inclusion of various thermal and zero temperature quantum corrections to the tree-level potential. 

Before moving on to details of this study, it is useful to review our current understanding of the EWPhT in singlet extensions of the SM. The strictly $Z_2$-preserving version of this model has been studied to great detail in Ref.~\cite{Espinosa:1993bs,Profumo:2007wc,Espinosa:2007qk,Espinosa:2011ax,Barger:2011vm,Curtin:2014jma,Kurup:2017dzf}, presented as the so-called "nightmare scenario'' for its challenges in testing it at future colliders. 
These scenarios generally enhance the EWPhT through loop effects of the singlet via the large quartic couplings (O(few)) between the singlet and  Higgs pairs. This, however, occurs in the regime where perturbative unitarity is in question, where the one-loop corrections are large, and further studies are in need. A special mechanism, where the EWPhT is enhanced by tree-level effects through a two-step phase transition, can also be realized in these scenarios \cite{Profumo:2007wc,Espinosa:2011ax,Barger:2011vm,Curtin:2014jma,Kurup:2017dzf}. However, once the requirement of a non-relativistic bubble wall motion is imposed, solutions under this category only exist in a narrow region of parameter space. For general $Z_2$-explicit breaking models, the large number of free parameters often requires  numerical studies which can provide benchmark point solutions \cite{Choi:1993cv,Ham:2004cf,Profumo:2007wc,Espinosa:2011ax,Barger:2011vm,Noble:2007kk,Ashoorioon:2009nf,Das:2009ue,Profumo:2014opa,Kotwal:2016tex}. The solutions in these scenarios often invoke additional tree-level barriers from the explicit $Z_2$ breaking terms.

For the well-motivated scenario we are considering, where the $Z_2$ is spontaneously broken, it is a priori not clear if a sufficiently strong first-order EWPhT can be in place.  First, a large mixing quartic coupling between the singlet pairs and the Higgs pairs is generically disfavored by Higgs precision tests, as this term will generate a sizable singlet-Higgs mixing when the singlet acquires a non-zero VEV. A small mixing quartic, instead, precludes a possible large loop effect from the singlet, which is one of the main mechanisms to enhance the EWPhT. Second, one might expect that the spontaneous $Z_2$ breaking singlet VEV could add additional trilinear terms and generate the $|H^\dagger H|^n$ or higher-order operators that could modify the Higgs potential directly via these tree-level couplings. Due to the relations among couplings in the spontaneous $Z_2$ breaking theory, it turns out that these operators are only generated at loop-level as if the $Z_2$ symmetry were not broken~\cite{Carena:2018vpt}. Hence this property prevents tree-level modifications to the Higgs potential that would be sizable enough to enhance the first-order EWPhT strength.

The above considerations imply that it is far from trivial to anticipate the behavior of the EWPhT in singlet SM extensions with spontaneously discrete symmetry breaking. Understanding the situation and the possible region of allowed parameter space for a strongly first-order EWPhT demands a detailed study, which is the purpose of this work.\footnote{ It is well known that domain wall problems are associated with the existence of multiple vacua in theories with spontaneous $Z_2$ breaking. However, domain wall problems can be alleviated by allowing for highly suppressed higher-dimensional operators that will minimally break the $Z_2$ symmetry explicitly.  Such highly suppressed contributions will not affect the discussion about phase transitions and their related phenomenology. We will not consider this issue any further in this work.}
As we shall show, we obtain a particular type of solutions that enhances the EWPhT via engineering nearly degenerate zero temperature vacua in a very predictive manner.
The paper is organized as follows: In Sec.~\ref{sec:theory} we introduce the SM extension and write down the expressions for the full one-loop potential and the daisy resummation. In Sec.~\ref{sec:EWPT} we classify the possible thermal histories, utilizing semi-analytic solutions that guide the understanding of our results. We also show the allowed region of parameter space for a strongly first-order EWPhT, and further check the robustness of our results against a nucleation calculation. An unavoidable,  distinctive feature of our study is the prediction of a light singlet-like scalar. We present the phenomenological consequences of this model studying the implications for the Higgs exotic decays, Higgs precision measurements, double Higgs production, and gravitational wave signatures in Sec.~\ref{sec:pheno}. Finally, we reserve Sec.~\ref{sec:summary} to conclude and Appendices A-D to show some specific details of our analysis.

\section{Singlet extension of the SM with spontaneous \boldmath{$Z_2$}-breaking }
\label{sec:theory}

\subsection{Tree-level potential}
\label{sec:tree}

We start with the tree-level Higgs boson  potential with an additional real singlet $s$:
\begin{equation} \label{}
\begin{split}
V_0 = - \mu_h^2 \phi^{\dagger} \phi + \lambda_h ( \phi^{\dagger} \phi )^2 + \frac{1}{2} \mu_s^2 s^2 + \frac{1}{4} \lambda_s s^4 + \frac{1}{2} \lambda_m s^2 (\phi^{\dagger} \phi) + V_{\rm SM}.
\end{split}
\end{equation}
There is an important discrete $Z_2$ symmetry in the singlet sector, under which $s\rightarrow -s$ and the rest of the fields remain unchanged. The singlet scalar field $s$ can spontaneously break this symmetry.

The SM Higgs doublet $\phi$ is written as 
\begin{equation} \label{}
\begin{split}
\phi = \frac{1}{\sqrt{2}} \begin{pmatrix} \chi_1+i\chi_2 \\ h+i\chi_3 \end{pmatrix},
\end{split}
\end{equation}
where $\chi_1$, $\chi_2$, $\chi_3$ are three Goldstone bosons, and $h$ is the Higgs boson. The tree-level potential of $h$ and $s$ in the unitary gauge reads
\begin{equation} \label{}
\begin{split}
V_0 (h, s) = -\frac{1}{2} \mu_h^2 h^2 + \frac{1}{4} \lambda_h h^4 + \frac{1}{2} \mu_s^2 s^2 + \frac{1}{4} \lambda_s s^4 + \frac{1}{4} \lambda_m s^2 h^2.
\end{split}
\end{equation}

At zero temperature, there are four non-degenerate extrema, with the possibility of the scalars having zero or non-zero VEVs. Amongst these four extrema, only two of them are consistent with the Higgs doublet obtaining a non-zero VEV. 
In this work, we are in particular interested in the case where the singlet also acquires a VEV.
The VEVs of the Higgs doublet and the real singlet in terms of the bare parameters of the potential can be written as
\begin{equation} \label{vev0}
\begin{split}
v|_{T=0} =v_{\rm EW}=\sqrt{\frac{2(2 \lambda_s \mu_h^2+\lambda_m \mu_s^2)}{4 \lambda_h \lambda_s - \lambda_m^2}} ,\quad w|_{T=0}= w_{\rm EW} =\sqrt{\frac{2(-2 \lambda_h \mu_s^2-\lambda_m \mu_h^2)}{4 \lambda_h \lambda_s - \lambda_m^2}} .
\end{split}
\end{equation}
The physical scalar masses are obtained by diagonalizing the squared mass matrix evaluated at the physical VEV,
\begin{equation} \label{mass0}
\begin{split}
M^2 &= \begin{pmatrix} \frac{\partial^2 V}{\partial h^2} &\frac{\partial^2 V}{\partial h \partial s} \\  \frac{\partial^2 V}{\partial h \partial s} & \frac{\partial^2 V}{\partial s^2}  \end{pmatrix}  \Big|_{(v_{\rm EW} ,w_{\rm EW} )}\\
&= \begin{pmatrix} 3h^2 \lambda_h - \mu_h^2+ \frac{1}{2} \lambda_m s^2  & \lambda_m hs \\  \lambda_m hs & \frac{1}{2} \lambda_m h^2  + \mu_s^2 + 3 \lambda_s s^2 \end{pmatrix} \Big|_{(v_{\rm EW} ,w_{\rm EW} )}.
\end{split}
\end{equation}

Electroweak Symmetry Breaking (EWSB) requires that the physical VEV $(v_{\rm EW}, w_{\rm EW})$ is the deepest minimum of the potential. For $(v_{\rm EW}, w_{\rm EW})$ to be a minimum,
\begin{equation} \label{eq:vstab}
\begin{split}
{\rm Det} M^2 = v_{\rm EW}^2 w_{\rm EW}^2 \left( 4 \lambda_h \lambda_s - \lambda_m^2\right) \ge 0,
\end{split}
\end{equation}
rendering  $4 \lambda_h \lambda_s - \lambda_m^2 \ge 0$ a necessary condition for  EWSB 
at tree level. 

There are five bare parameters $\{\mu_h^2, \mu_s^2, \lambda_h, \lambda_s, \lambda_m\}$ in the tree-level potential. They can be traded by five physical parameters, two of which, the Higgs VEV and the Higgs mass $m_H$, are fixed by boundary conditions  
\begin{equation} 
\begin{split}
v_{\rm EW}= 246\ {\rm GeV},\quad m_H=125\ {\rm GeV}.
\end{split}
\end{equation}
The remaining three physical parameters are related to the singlet VEV, the singlet mass and the mixing angle of the mass eigenstates, and we defined $\tan \beta = w_{\rm EW} /v_{\rm EW}$. Detailed discussion of the parametrization can be found in Appendix~\ref{sec:physpara}.

\subsection{One-loop effective potential at finite temperature}
\label{sec:thermal}

The one-loop effective potential at finite temperatures is calculated in the background of the Higgs and singlet fields. Effective masses of all degrees of freedom in the plasma dependent on the background fields are: 
\begin{equation} \label{}
\begin{split}
&m_W^2 (h,s) = \frac{g^2}{4} h^2,\quad m_Z^2 (h,s)= \frac{g^{'2}+g^2}{4} h^2,\quad m_t^2 (h,s) = \frac{1}{2} h_t^2 h^2,\\
&m_{\chi_{1,2,3}}^2 (h,s) =-\mu_h^2+ \lambda_h h^2 + \frac{1}{2} \lambda_m s^2,\\
&m_h^2 (h,s)=  -\mu_h^2 +  3 \lambda_h h^2 + \frac{1}{2} \lambda_m s^2,\\
&m_s^2 (h,s)= \mu_s^2 + \frac{1}{2} \lambda_m h^2+3 \lambda_s s^2,\\
&m_{sh}^2 (h,s)=\lambda_m h s,\\
\end{split}
\end{equation}
where $\chi_{1,2,3} $  are the Goldstone bosons and the particle degrees of freedom are: 
\begin{equation} \label{}
\begin{split}
n_W=6,\ n_Z=3,\ n_t=-12,\ n_h=1,\ n_{\chi_{1,2,3}}=1,\ n_s=1.
\end{split}
\end{equation}
For the Higgs and singlet degrees of freedom, mass eigenvalues entering the effective potential are
\begin{equation} \label{egm}
\begin{split}
m_{\varphi_1, \varphi_2}^2 (h,s)= \frac{1}{2} \Big\{ &( 3 \lambda_h + \lambda_m/2) h^2 + (3 \lambda_s + \lambda_m/2) s^2 - \mu_h^2 + \mu_s^2\\
 &\pm \sqrt{\big[( 3 \lambda_h - \lambda_m/2) h^2 + (-3 \lambda_s + \lambda_m/2) s^2 - \mu_h^2 - \mu_s^2\big]^2 + 4 \lambda_m^2 s^2 h^2}  \Big\},
\end{split}
\end{equation}
where $\varphi_1, \varphi_2$ are  the Higgs and singlet mass eigenstates  with particle degrees of freedom $n_{\varphi_1, \varphi_2}=1$.

In this study we work in the Landau gauge and the Goldstone modes contribute separately in addition to the massive bosons. There has been ample discussion in the literature on the issue of gauge dependence  in 
 perturbative calculations  of the effective potential, both at zero and finite temperature \cite{Jackiw:1974cv,Kang:1974yj,Dolan:1974gu,Fukuda:1975di,Aitchison:1983ns,Loinaz:1997td,Patel:2011th,Garny:2012cg,Andreassen:2014eha,Andreassen:2014gha}. In that sense, we understand that our treatment is not manifestly gauge invariant. We expect, however, that  our analysis provides a realistic
  estimate of the EWPhT strength.\footnote{The reason for this is that gauge dependence appears at loop level in perturbation theory, while,  as will be discussed later in the paper, in our model the important enhancement of the EWPhT strength, $v_c/T_c > 1$,  is due to tree level effects in the potential that come into play once the finite temperature barrier turns on. Indeed, as we  will discuss in Section \ref{sec:EWPT},
the  thermal contributions are subdominant. We however intend to study the effects of gauge dependence further in a future work.}

The temperature dependent part of the one-loop effective potential  \cite{Quiros:1994dr}, referred in the following as (one-loop) thermal potential, reads
\begin{equation} \label{}
\begin{split}
V^T_{\rm 1-loop} (h, s, T)
&= \frac{T^4}{2\pi^2} \left[ \sum_{B} n_B J_B \left(\frac{m_B^2(h,s)} {T^2}\right) + \sum_{F} n_F J_F \left(\frac{m_F^2(h,s)} {T^2}\right)  \right],
\end{split}
\end{equation}
where $B$ includes all the bosonic degrees of freedom that couple directly to the Higgs boson, namely  $W, Z, \chi_i, \varphi_1, \varphi_2 $, and $F$ stands for the top quark fermion only.
The $J_B$ and $J_F$ functions for bosons and fermions 
can be evaluated by numerical integration or proper extrapolation. All the numerical study in this work is performed using a modified version of {\tt CosmoTransitions}~\cite{Wainwright:2011kj}, where spline interpolation is implemented. The spline interpolation shows the best agreement with our full numerical integration results. 
For better analytical control, we use high-temperature expansion for analytical analyses in the next section. Appendix~\ref{sec:VT} provides a more detailed discussion of the thermal potential, including formalism of the $J$ functions, numerical convergence, and the high-temperature expansion.
 
The Coleman-Weinberg (CW) potential \cite{Coleman:1973jx} is the temperature-independent part of the effective potential at one-loop order
\bea
V_{\rm CW} (h,s)
&= \frac{1}{64\pi^2} \left(  \sum_{B}n_B m_B^4(h,s) \Big[ \log \Big( \frac{m_B^2(h,s)}{Q^2} \Big) - c_B \Big]\right. \nonumber \\
&\left. -  \sum_{F}  n_F m_F^4(h, s) \Big[ \log \Big( \frac{m_F^2(h,s)}{Q^2} \Big) - \frac{3}{2} \Big] \right),
\eea
where $B$ and  $F$ were defined above and  $c_B = 3/2 (5/6) $ for scalar (vector) bosons. The potential is calculated in the dimensional regularization and the $\overline{\rm MS}$ renormalization scheme. Counterterms have been added to remove the UV divergences. $Q$ is the renormalization scale that we have chosen to $Q=1000$~GeV (see discussion in Appendix~\ref{sec:scheme}). This part of the one-loop effective potential gives corrections to both the Higgs VEV and the Higgs mass at zero temperature; hence, the bare parameters deviate from their tree-level values to satisfy the boundary conditions (for more details see Appendix~\ref{sec:scheme}). 
In our numerical studies, we perform a 5-dimensional scan of the bare model parameters, selecting those consistent with the SM Higgs VEV $v_{\rm EW}\simeq 246$~GeV and the Higgs-like particle mass $m_{\varphi_i}$ $\simeq 125$~GeV, with $i = 1$ or 2 depending on the mass hierarchy between mass eigenstates, where we allow for an uncertainty of  $\pm$2~GeV in the VEV and the mass value, respectively.  Observe that adding the CW contributions is required to perform a consistent one-loop calculation,  but significantly decreases the efficiency of the numerical scanning in comparison to  the only one-loop thermal potential approximation, for which the number of scanning parameters is reduced to three. 

Lastly,  corrections from daisy resummation of ring diagrams need to be included in the full one-loop potential to ensure validity of the perturbative expansion. The leading order resummation results give thermal corrections of $\Pi_i = d_i T^2$ to effective masses, say $m_i^2(h, s) \to m_i^2(h,s, T) =  m_i^2(h, s) + d_i T^2$, where $d_i$ for different degrees of freedom in the plasma are~\cite{Curtin:2014jma}
\begin{equation} \label{}
\begin{split}
&{ d_{W^{\pm,3}}^L} = \frac{11}{6} g^2,\quad d_{W^{\pm,3}}^T = 0,\quad { d_B^L = \frac{11}{6} g^{'2} },\quad d_b^T = 0,\\
&d_{\chi} ={ \frac{3}{16} g^2 + \frac{1}{16} g^{'2}  } + \frac{1}{2} \lambda_h + \frac{1}{4} y_t^2 + \frac{1}{24} \lambda_m,\\
&d_{hh} ={ \frac{3}{16} g^2 + \frac{1}{16} g^{'2}  } + { \frac{1}{2} } \lambda_h + \frac{1}{4} y_t^2 + \frac{1}{24} \lambda_m,\quad 
d_{ss} =\frac{1}{4} \lambda_s + \frac{1}{6} \lambda_m,\quad 
d_{sh} \approx 0.
\end{split}
\end{equation}
A truncated full dressing implementation corresponds to replacing all $m_i^2(h, s)$ with $m_i^2(h,s, T)$ in the one-loop effective potential at finite temperatures~\cite{Arnold:1992rz,Curtin:2016urg}.


\section{Enhancing the Electroweak phase transitions}
\label{sec:EWPT}

\begin{figure}
\centering
\begin{subfigure}{0.48\textwidth}
\centering
\includegraphics[width =\textwidth]{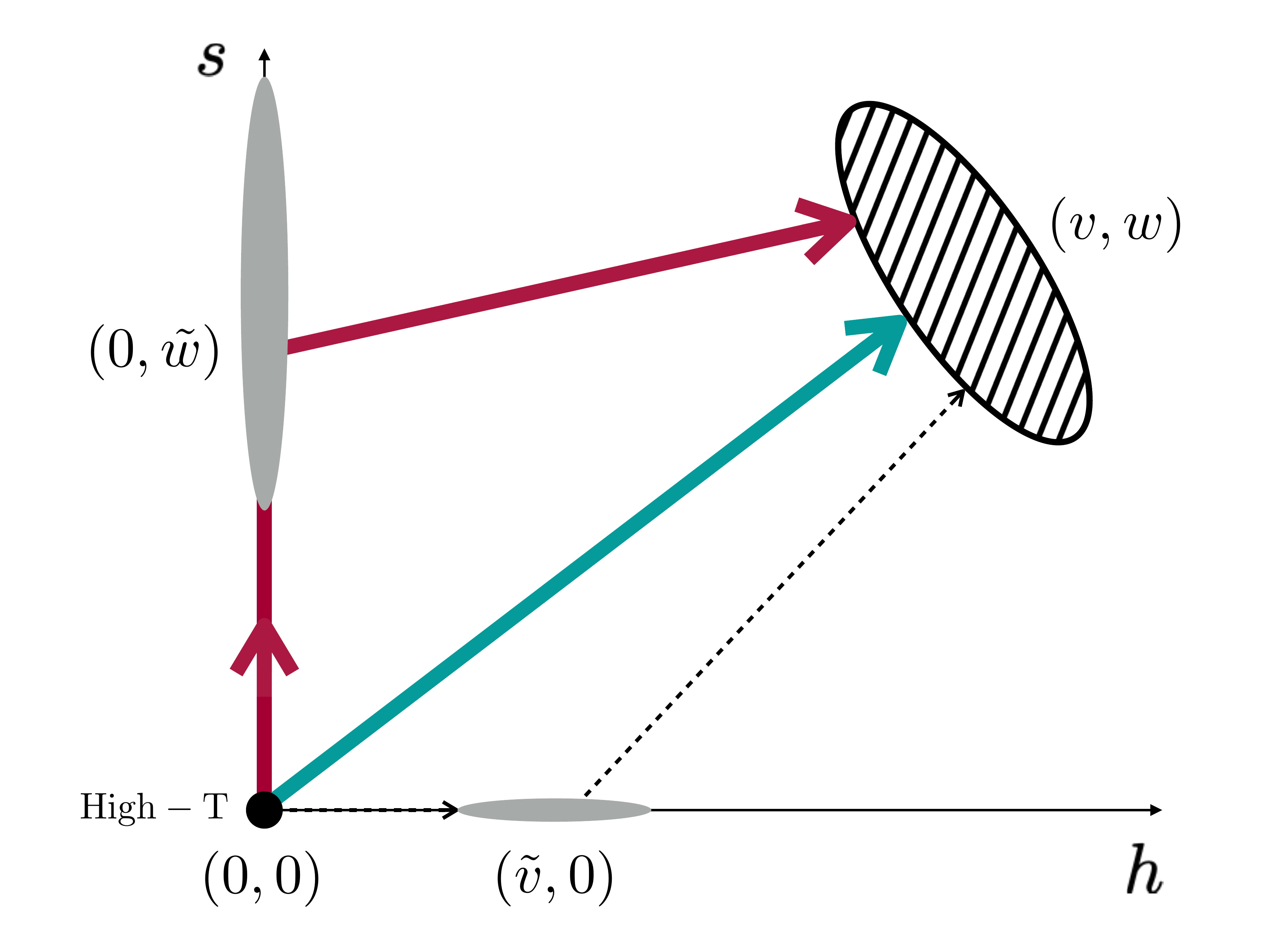}
\end{subfigure}
\begin{subfigure}{0.48\textwidth}
\centering
\includegraphics[width =\textwidth]{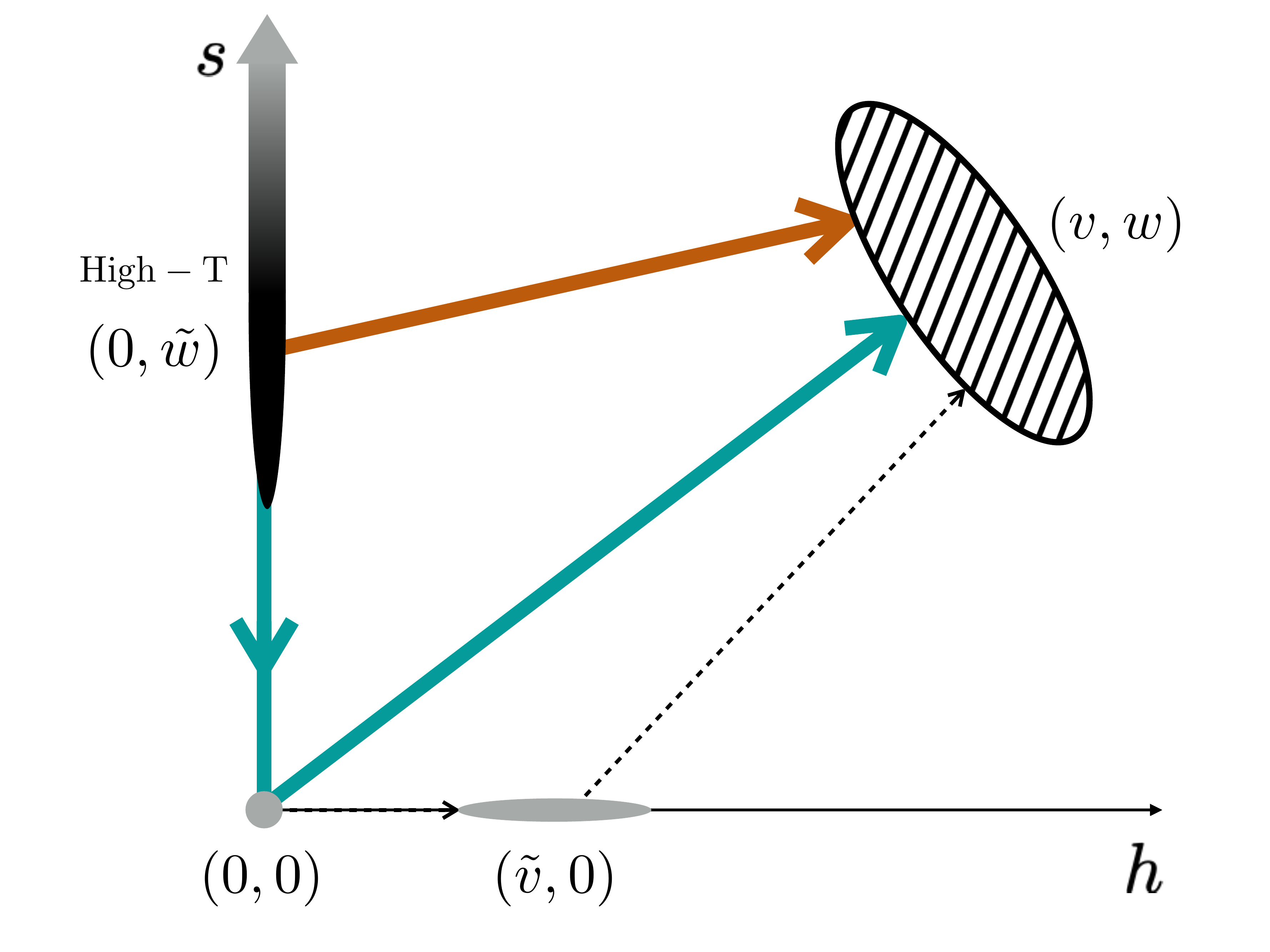}
\end{subfigure}
\caption{Schematic picture of the thermal histories with different phase transition patterns. Left: restoration scenarios, where the thermal history starts from the symmetric phase at $(0,0)$. Right: non-restoration scenarios, where the thermal history starts from the $Z_2$ non-restored phase at $(0,\tilde{w})$. Phases are represented by bubble areas: the high temperature phase by a black bubble, the zero temperature phase by a hatched bubble, and the intermedia phases by gray bubbles. Phase transition steps are represented by arrow lines with a color code depicting the different scenarios: {\color{magenta} scenario A}, {\color{orange} scenario A-NR}, {\color{teal} scenario B/B-NR}, respectively. The dashed arrow line scenario is discussed in Appendix~\ref{sec:exotics}.}
\label{fig:pattern}
\end{figure}

In this section, we analyze all possible electroweak phase transition patterns appearing in our real singlet scalar extension of the SM. The thermal history could be very rich, as depicted in Fig.~\ref{fig:pattern}. We highlight the cases for which a strongly first-order electroweak phase transition that is consistent with current SM EW and Higgs precision data is feasible.

Before proceeding with a more detailed analysis, we shall briefly described the possible  thermal histories  for  the scalar potential defined in the previous section. 
The spontaneous $Z_2$ breaking singlet extension of the SM differs from the $Z_2$-preserving case significantly through the allowed size of the mixing quartic coupling $\lambda_m$. Large $\lambda_m$ certainly helps with enhancing the EWPhT by enhancing the thermal barrier term $E T h^3$ since the singlet is a new bosonic degree of freedom. However, in the spontaneous $Z_2$ breaking case, $\lambda_m$ is not an independent free parameter, but rather proportional to the Singlet-Higgs mixing angle $\sin\theta$, which in turn is constrained by  LHC Higgs precision data to be smaller than 0.4.
Hence, in the spontaneous $Z_2$ breaking case, the smaller size of $\lambda_m$ 
 implies that a sufficiently strong first-order  EWPhT is only achievable via more subtle effects in the potential.

For scenarios of our interests, both the electroweak symmetry and the $Z_2$ symmetry are broken at zero temperature. At high temperatures, instead, the electroweak symmetry is preserved (high-temperature restoration of the EW symmetry), and the $Z_2$ symmetry can be either broken or restored. As a result, we will show how the path to the zero temperature electroweak physical vacuum can involve a one- or two-step phase transition.


In the following, we shall focus on the following four relevant scenarios:
\begin{itemize}
  \item {\color{blue} Scenario A}: Two-step phase transition
  \item[]\begin{center}{\color{blue} (0,0)$\to$(0,$\tilde w$)$\to$($v$,$w$)}\end{center}
  \item {\color{blue} Scenario B}: One-step phase transition
  \item[] \begin{center}{\color{blue} (0,0)$\to$($v$,$w$)}\end{center}
\end{itemize}
and their corresponding counterparts with $Z_2$ non-restoration (NR) at high temperatures:
\begin{itemize}
  \item {\color{blue} Scenario A-NR}: One-step phase transition
  \item[]\begin{center}{\color{blue} (0,$\tilde w$)$\to$($v$,$w$)}\end{center}
  \item {\color{blue} Scenario B-NR}: Two-step phase transition
  \item[] \begin{center}{\color{blue} (0,$\tilde w$)$\to$(0,0)$\to$($v$,$w$)}\end{center}
\end{itemize}
The correspondence between the restoration and non-restoration scenarios  is defined by  them sharing the same final path towards the electroweak physical vacuum.  All the minima defined above are temperature dependent, and different VEVs are associated with  different paths in the thermal history. 

There exist other possible scenarios, in which although the final step towards the true EW vacuum can involve a strongly first-order phase transition, it occurs when the sphalerons are already inactive.  In such cases, the temperature at which the sphalerons are still active is associated with a previous step in which the EW symmetry breaking yields a false EW breaking vacuum and does not involve a sufficiently strongly first-order phase transition. 
These scenarios are:
\begin{itemize}
  \item[]\begin{center}{(0,0)$\to$($\tilde v$,0)$\to$($v$,$w$)}\end{center}
    \item[]\begin{center}{(0,$\tilde w$)$\to$(0,0)$\to$($\tilde v$,0)$\to$($v$,$w$)}\end{center}
\end{itemize}
For completeness, they are  briefly discussed  in Appendix~\ref{sec:exotics}, however, they are not of interests to our study.

\subsection{Scenario A: (0,0)$\to$(0,$\tilde w$)$\to$($v$,$w$)}
\label{sec:SA}

We shall show that the electroweak phase transition can be strongly first-order if the transition occurs from a $Z_2$  breaking/EW preserving vacuum, $(0, \tilde{w})$, to the true EW physical vacuum with $Z_2$  breaking, $(v, w)$.
This behavior can develop in two different ways: the one discussed in this subsection, scenario A,  that involves a two-step transition  in which at high temperatures the system is in a symmetric vacuum $(0, 0)$, and then evolves to a  spontaneous  $Z_2$  breaking/EW preserving  vacuum at lower temperatures, to final transition to the true EW physical vacuum with  $Z_2$  breaking.
A different, one step phase transition path, that we call scenario A-NR, in which the system starts directly at a  $Z_2$  breaking/EW preserving  vacuum  at high temperatures and then transitions to  the true EW physical vacuum with  $Z_2$  breaking, will be discussed in detail in Sec.~\ref{sec:nR}. 

First, we start considering a high-temperature expansion to show analytically the  behavior. Under the high-temperature expansion, the finite temperature potential (without CW potential and daisy resummation) is given in Eq.~(\ref{eq:htv}). The complicated field-dependent term $-E(h, s)T$ 
can enhance the trilinear coefficient E beyond the SM value used in Eq.~(\ref{eq:vap}) below, due to the effect of the additional quartic couplings. For simplicity, however,  we shall neglect such subdominant effects in the following analytical considerations.
Without such a term, the effective potential reads
 \begin{equation} \label{eq:vap}
\begin{split}
V(h,s, T) \approx \frac{1}{2}(-\mu_h^2 + c_hT^2) h^2  - E^{\rm SM}T h^3 + \frac{1}{4} \lambda_h h^4+ \frac{1}{2}( \mu_s^2 + c_s T^2) s^2 + \frac{1}{4} \lambda_s s^4 + \frac{1}{4} \lambda_m s^2h^2 ,
\end{split}
\end{equation}
where relevant coefficients are given in the Appendix~\ref{sec:VT}.  

For scenario A, the electroweak symmetry breaking proceeds through the second step from a $Z_2$  breaking/EW preserving vacuum, $(0, \tilde{w})$, to the true EW physical vacuum with $Z_2$  breaking, $(v, w)$, at a critical temperature $T_c$ given by
\begin{equation} \label{eq:tc}
\begin{split}
T_c^2 = \frac{2 \lambda_s \mu_h^2+\lambda_m \mu_s^2 }{2c_h \lambda_s- c_s \lambda_m - 16 \frac{(E^{\rm SM})^2 \lambda_s^2}{4 \lambda_h \lambda_s - \lambda_m^2} },
\end{split}
\end{equation}
where both vacua coexist and are degenerate. In the  $Z_2$  breaking/EW preserving  vacuum, the singlet has a temperature dependent VEV that  at
 $T_c$ reads
\begin{equation}
\begin{split}
\tilde{w} (T_c) = \sqrt{\frac{-\mu_s^2 - c_s T_c^2}{\lambda_s}} ,
\end{split}
\end{equation}
while in the true EW physical vacuum with $Z_2$  breaking, both the Higgs and the singlet fields have non-zero temperature dependent VEVs which at
 $T_c$  respectively read

\begin{equation}
\begin{split}
&
v_c \equiv v(T_c) =  \frac{8 E^{\rm SM} \lambda_s}{4 \lambda_h \lambda_s - \lambda_m^2} T_c ,\;\;\;\;\; \;\;\; w(T_c) = \sqrt{\frac{-\mu_s^2}{\lambda_s}-T_c^2  \big[\frac{c_s}{\lambda_s} + 32 \frac{ (E^{\rm SM})^2\lambda_s \lambda_m} {4 \lambda_h \lambda_s-\lambda_m^2} \big] } .
\end{split}
\end{equation}
The phase transition strength is determined by the ratio
\begin{equation} \label{eq:a}
\begin{split}
\frac{v_c}{T_c}= \frac{2 E^{\rm SM}}{ \lambda_h - \lambda_m^2/(4 \lambda_s)} = \frac{2 E^{\rm SM}}{\lambda_{h}^{\rm SM}} \Big[1 + \sin^2\theta \frac{m_H^2 -m_S^2}{m_S^2} \Big],
\end{split}
\end{equation}
where a sufficiently strong first-order phase transition requires $\frac{v_c}{T_c} \gtrsim 1$.
Accordingly, the EWPhT strength can be enhanced by having smaller singlet scalar mass $m_S$ compared to the Higgs boson  mass $m_H$. The lighter the singlet scalar and the larger the Higgs-singlet mixing parameter, $\sin\theta$,  the stronger the phase transition. In terms of the bare parameters, we observe that the strength of the phase transition is governed by the magnitude of the  effective quartic coupling defined as
\begin{equation}
\begin{split}
\tilde{\lambda}_h \equiv \lambda_h - \lambda_m^2/(4 \lambda_s)\end{split}.
\end{equation}
Notice that the EWSB condition  shown in Eq.~(\ref{eq:vstab}) requires  $\tilde{\lambda}_h \le 0$,
which ensures $\frac{v_c}{T_c}$ being positive definite, without constraining its absolute value.
  $\tilde{\lambda}_h  \gtrsim 0$ is the near criticality condition for EWSB, which at the same time yields  maximal  enhancement  of the strength of the EWPhT.

\begin{figure}
\centering
\includegraphics[width =0.8\textwidth]{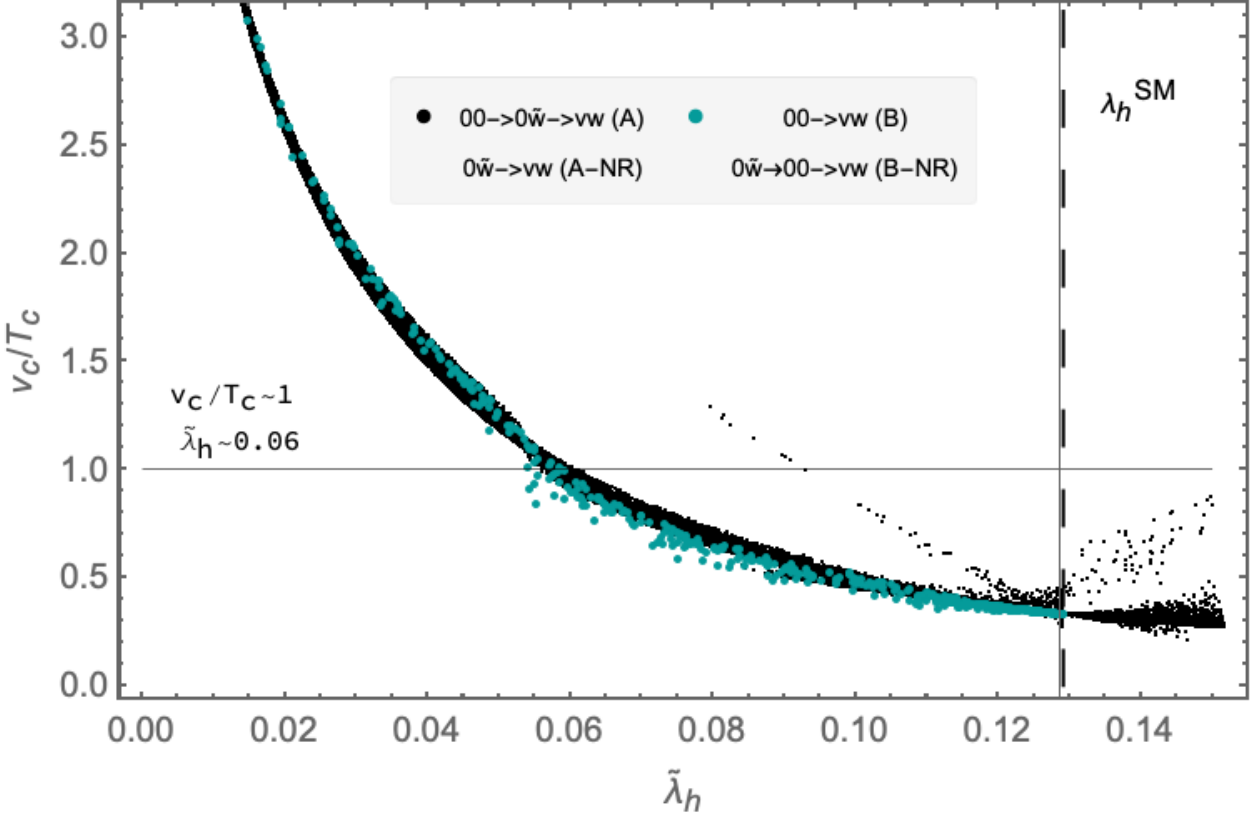}
\includegraphics[width =0.8\textwidth]{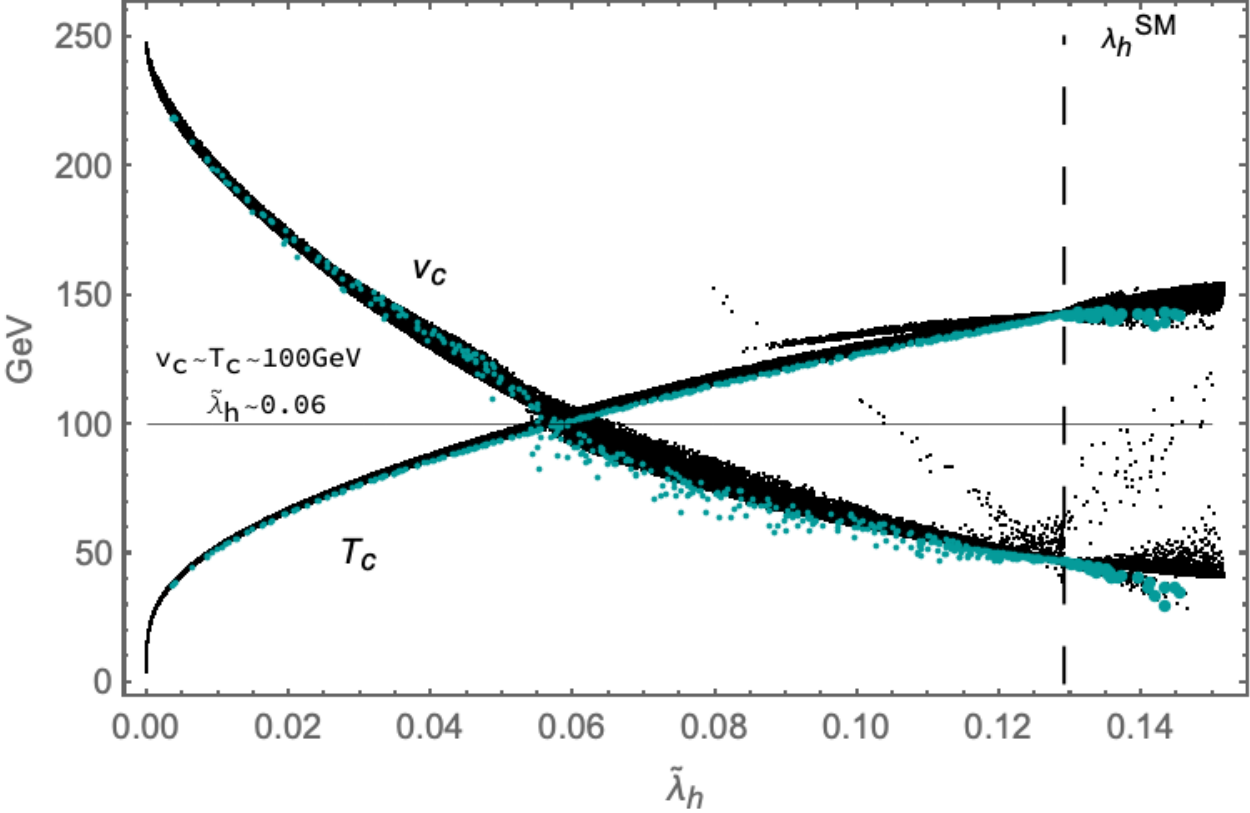}
\caption{Results for the electroweak phase transition in a spontaneous $Z_2$ breaking singlet extension of the SM, with full numerical study of the one-loop thermal potential. EWPhT information of scenario A and A-NR are shown in black dots and scenario B and B-NR are shown in green dots. Upper panel: $v_c/T_c$ versus the effective quartic coupling $\tilde \lambda_h$. Lower panel: $v_c$ and $T_c$ versus $\tilde \lambda_h$.}
\label{fig:pt}
\end{figure}

In the following we discuss the behavior of the potential at zero temperature, that will provide information of the potential energy difference between the true EW physical vacuum and the  $Z_2$  breaking/EW preserving extremum at zero temperature, which in turn has information on the magnitude of the critical temperature, and hence on the strength of the EWPhT.  Moreover, to better understand the EWPhT behavior, we shall further discuss the dependence of the relevant quantities at the critical temperature in terms of the model parameter $\tilde \lambda_h$ that governs them.

At zero temperature, the tree-level potential difference between the true vacuum, $ (v_{\rm EW}, w_{\rm EW})$, and the $Z_2$  breaking/EW preserving extremum, $(0, \tilde{w}|_{T=0})$ is given by,
\begin{equation}
\begin{split}
\Delta V_{\rm A}  &\equiv V(0, \tilde{w}|_{T=0}, T=0) - V(v_{\rm EW}, w_{\rm EW}, T=0) = \frac{v^4}{4} \left( \lambda_h - \frac{\lambda_m^2}{4 \lambda_s}\right) = \frac{v^4}{4} \tilde{\lambda}_h.
\end{split}
\label{eq:tildelh}
\end{equation}
This zero temperature potential  energy difference reduces to  the SM value, $\Delta V^{\rm SM} = V^{\rm SM}(0) - V^{\rm SM}(v_{\rm EW}) = \frac{v_{\rm EW}^4}{4}  \lambda_h^{\rm SM}$, in the limit in which the singlet decouples. Eq.(\ref{eq:tildelh}) depicts the proportionality between  $\Delta V_{\rm A}$ and $ \tilde{\lambda}_h$, and implies that near criticality, for which $\tilde{\lambda}_h$ is small, 
 $\Delta V_{\rm A}$ is small  as well. 
 Given Eq.~(\ref{eq:a}), we see that a small value of  $\Delta V_{\rm A}$ is naturally associated to a large value of $\frac{v_c}{T_c}$.


When the $Z_2$  breaking/EW preserving extremum and the true vacuum have less potential energy difference at zero temperature, the critical temperature is lower.
 We consider now the specific dependence of the critical temperature on the model parameter  $\tilde{\lambda}_h$.
The thermal evolution of the  two zero temperature extrema  is controlled by temperature dependent coefficients in the thermal potential. More specifically, we rewrite the critical temperature in Eq.~(\ref{eq:tc}) as
\begin{equation}
\begin{split}
T_c^2 = v^2 \frac{  \tilde{\lambda}_h^2}{\left( c_h - \frac{\lambda_m}{2\lambda_s} c_s\right)   \tilde{\lambda}_h - 2 ( E^{\rm SM} )^2},
\end{split}
\end{equation}
where $( E^{\rm SM} )^2 \sim 10^{-4}$ and $ c_h - \frac{\lambda_m}{2\lambda_s} c_s \approx 0.33 + \frac{1}{2} \lambda_h - \frac{\lambda_m}{12}\left( 1+ \frac{\lambda_m}{\lambda_s} \right)$. Numerically, the $( E^{\rm SM} )^2$ term is negligible and we shall  drop it. This corresponds to the fact that the temperature dependent quadratic terms dominate the thermal evolution. The critical temperature then reads
\begin{equation}
\begin{split}
T_c \simeq  \frac{v}{\sqrt{c_h - \frac{\lambda_m}{2\lambda_s} c_s}}  \tilde{\lambda}_h^{\frac{1}{2}}.
\end{split}
\end{equation}
We observe that near criticality,
the critical temperature is very close to zero. Meanwhile, the Higgs VEV at the critical temperature is larger and closer to the zero temperature VEV of 246~GeV. 
More specifically,
\begin{equation}
\begin{split}
v_c = \frac{2 E^{\rm SM}}{\lambda_h - \frac{\lambda_m^2}{4 \lambda_s}} T_c \simeq 2 E^{\rm SM}  \frac{v}{\sqrt{c_h - \frac{\lambda_m}{2\lambda_s} c_s}} \tilde{\lambda}_h^{-\frac{1}{2}}.
\end{split}
\end{equation}
Notice that the $E^{\rm SM}$ factor here, or else the trilinear term in the thermal potential, is required to give a non-zero value of $v_c$. $E^{\rm SM}$ is not essential to render a low critical temperature, 
but does ensure that the phase transition is first-order instead of second-order. 

In summary, we have determined all relevant quantities to  the phase transition strength at the critical temperature in terms of  the effective quartic coupling $\tilde{\lambda}_h$, that controls  our model behavior, as
\begin{equation} \label{eq:ltscale}
\begin{split}
\Delta V_{\rm A} \propto \tilde{\lambda}_h,\quad T_c \propto \tilde{\lambda}_h^{\frac{1}{2}},\quad v_c \propto \tilde{\lambda}_h^{-\frac{1}{2}},\quad \frac{v_c}{T_c}  \propto \tilde{\lambda}_h^{-1}.
\end{split}
\end{equation}
{\it Within the mean field analysis considered,} the effective quartic coupling $\tilde{\lambda}_h$ is bounded from above 
by the Higgs quartic coupling $\lambda_h$
, and from below at $0$ by EWSB requirements. 
The near criticality condition, 
which corresponds to small values of $\tilde{\lambda}_h$, yields 
low values of the critical temperature and, therefore, a SFOPhT.

Fig.~\ref{fig:pt} shows numerical results obtained with {\tt CosmoTransitions} with full consideration of the one-loop thermal potential, as shown by the scattered black points.
The dependence of $v_c$, $T_c$, and the transition strength $v_c/T_c$ on the effective quartic coupling $\tilde{\lambda}_h$, shows excellent  agreement with our analytical results \footnote{The agreement is excellent in the low $\tilde{\lambda}_h$ region, while for larger values of  $\tilde{\lambda}_h$ other effects, for example those from  thermal trilinear terms, start to contribute and dominate over the tree-level effect associated with small $\tilde{\lambda}_h$. Such effects could possibly enhance the EWPhT; however, we did not find a relevant enhancement. Thus we do not further discuss them in the remaining of this work}
derived within a high-temperature expansion of the one-loop thermal potential, as shown in Eq.~(\ref{eq:ltscale}).
Fig.~\ref{fig:pt} also includes results for other scenarios that will be discussed below.

The enhancement of the phase transition strength due to the reduction of the potential depth at zero temperature has been discussed in the literature in other contexts triggered by loop effects \cite{Espinosa:2007qk,Curtin:2014jma,Kurup:2017dzf,Harman:2015gif}. However, when such a sizable reduction of the potential depth is due to loop effects, it requires sizable couplings, which in turn may break perturbativity, or it needs multiple singlets. 
In our scenarios, the potential depth reduction at zero temperature arises at tree level, similar to some other SM extensions \cite{Huang:2014ifa,Harman:2015gif,Dorsch:2017nza}, and relies on the spontaneous breaking of the $Z_2$ symmetry. These effects could be sizable even for sufficiently small coupling constants, which open a window to interesting  Higgs phenomenology.


In Fig.~\ref{fig:para}, we show the same data set from the numerical scan as in Fig.~\ref{fig:pt}, but 
depicted in the $c_s-\mu_s^2$ plane of model parameters, where 
$ c_s \equiv  \frac{1}{12} (2\lambda_m + 3 \lambda_s  )$ is a parameter controlling the boundary between high temperature $Z_2$ restoration and non-restoration behaviors, as will be discussed in detail in  Sec.~\ref{sec:nR}. Scenario A 
is shown in burgundy, and 
regions rendering SFOPhT are shown with a burgundy  darker shade.  
In this figure, we also show the approximated boundaries for SFOPhT, in burgundy solid  and dashed  lines, that are obtained from the mean field analysis with $\tilde{\lambda}_h \sim 0.06$, that  is the value of $\tilde{\lambda}_h$ at which $v_c /T_c \approx 1$, as  obtained from numerical estimation (see  Fig.~\ref{fig:pt}). The contours agree well with the dark region of SFOPhT from the numerical scanning. 
Points inside the burgundy solid and dashed lines are for values of $\tilde{\lambda}_h \lesssim 0.06$ as required for SFOPhT.
 We shall discuss this figure in further detail when considering  the other scenarios, including those with non-restoration of the  $Z_2$ symmetry.
\subsection{Scenario B: (0,0)$\to$($v$,$w$)}
\label{sec:SB}


A direct one-step phase transition from a fully symmetric phase to the physical vacuum could be realized in restricted regions of parameter space, while allowing for a strong first-order EWPhT. As we shall discuss in the following, such a one-step transition requires a comparable critical temperature for the $(0, 0)\to (\tilde{v}, 0)$ and $(0, 0)\to (0, \tilde {w})$.

\begin{figure}
\centering
\includegraphics[width =0.8\textwidth]{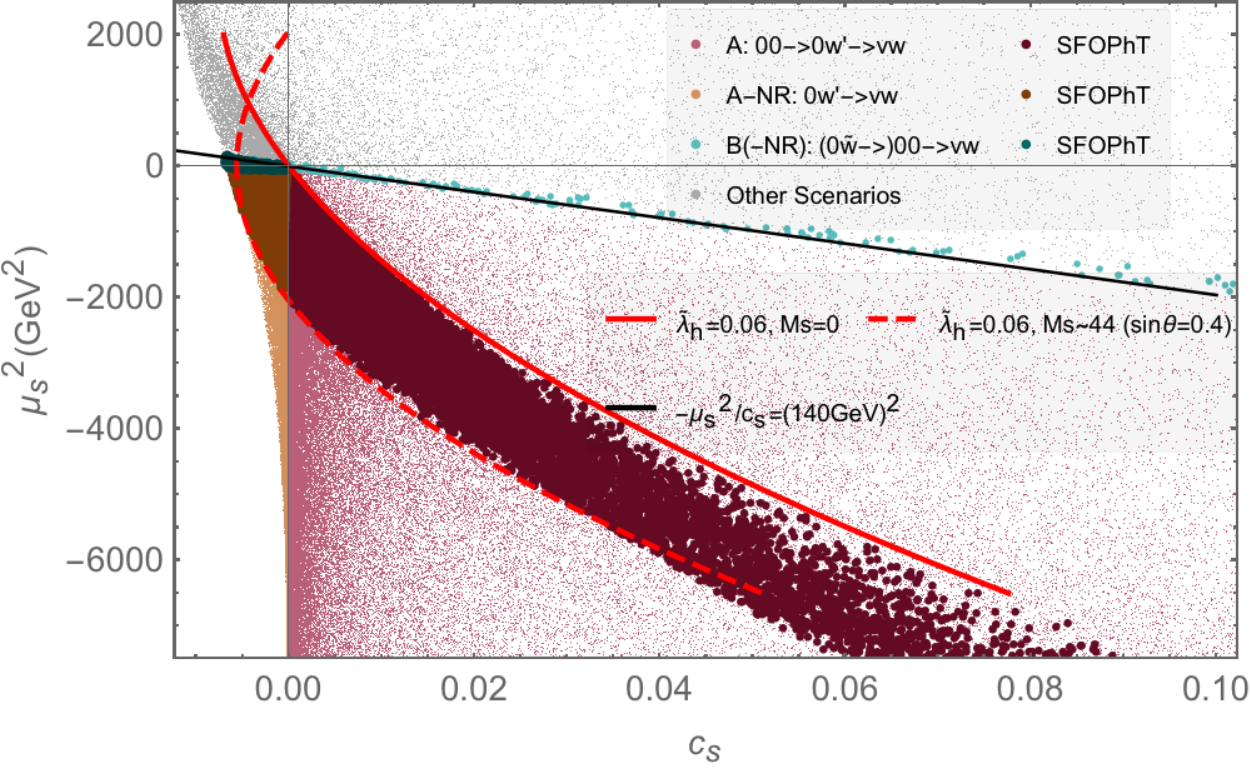}
\caption{Parameter space on the $c_s$-$\mu_s^2$ plane with different phase transition scenarios. Color scheme of the scattered points for different scenarios: {\color{magenta} scenario A}: two step phase transition with $c_s\ge 0$; {\color{orange} scenario A-NR}: one step phase transition with $c_s < 0$; {\color{teal} scenario B/B-NR}: one/two step phase transition with positive/negative $c_s$. Darker regions correspond to regions rendering strong first-order electroweak phase transitions for specific scenarios. Rough boundaries of $\tilde{\lambda}_h \sim 0.06$ for strong first-order EWPhT are shown.
The solid red boundary is a boundary under the limit of $m_S \to 0$ when $\tilde{\lambda}_h \sim 0.06$. The dashed red boundary is a boundary at $\sin\theta = 0.4$ (corresponds to $m_S \approx 44$ GeV provided $\tilde{\lambda}_h \sim 0.06$).  Points inside the burgundy solid and dashed lines are for values of $\tilde{\lambda}_h \lesssim 0.06$ as required for SFOPhT (corresponding to nearly degenerate minima at zero temperature). The fine tuned region for scenario B is featured by the condition $\frac{-\mu_s^2}{c_s} \sim (140$ GeV$)^2$ (shown in black line).}
\label{fig:para}
\end{figure}

In Fig.~\ref{fig:para}, we show in green a scan of points for
%
 scenario B (and its non-restoration counterpart, scenario B-NR to be discussed later on), whereas  regions rendering SFOPhT are shown in a darker green shade. As we observe in  Fig.~\ref{fig:para}, the scenario B lies within a narrow restricted region where $\sqrt{\frac{-\mu_s^2}{c_s}} \sim 140$ GeV (shown as a black line in the figure). This can be understood in the sense that $\sqrt{\frac{-\mu_s^2}{c_s}} $ features the temperature of $Z_2$ breaking, while $140$ GeV features the temperature of the electroweak breaking in the limit of decoupling the singlet. When these two temperatures are comparable, the $Z_2$ symmetry and the electroweak symmetry may break simultaneously, which is realized in scenario B through the phase transition step $(0, 0) \to (v, w)$. Observe that, given our knowledge of EWPhT in the SM,  we would need the actual temperature of simultaneous $Z_2$/EW breaking to be below $\sqrt{\frac{-\mu_s^2}{c_s}} \sim 140$ GeV, if we expect this scenario to allow for a sufficiently strong first-order EWPhT. We shall study this in the following.

Using the high temperature expansion of the effective potential, Eq.(\ref{eq:vap}),  we can compute analytically, for  scenario B (and similarly for scenario B-NR),  the  strength of the phase transition by solving for the ratio,
\begin{equation} \label{eq:1step}
\begin{split}
 \frac{v_c}{T_c} = \frac{2 E^{\rm SM}}{\tilde{\lambda}_h  + \frac{(\mu_s^2/T_c^2 + c_s)^2}{ \lambda_s\left[\frac{v(T_c)}{T_c}\right]^4}}.
\end{split}
\end{equation}
In the above,  $\tilde{\lambda}_h$ is defined as in Eq.~(\ref{eq:tildelh}) and $T_c$ is the critical temperature at which the $Z_2$/EW symmetric vacuum, $(0, 0)$, is  degenerate with the physical vacuum, $(v, w)$. 
Since both terms in the denominator are positive definite (without one-loop Coleman-Weinberg correction), they must be sufficiently small for the transition to be strongly first-order.
Indeed, the second term in the denominator, $(\mu_s^2/T_c^2 + c_s)^2/(\lambda_s\left[\frac{v(T_c)}{T_c}\right]^4)$, is numerically small for scenario B, and 
%
one can then approximate the $v_c / T_c $ ratio by 
\begin{equation} \label{eq:1stepaprox}
\begin{split}
\frac{v_c}{T_c} \simeq \frac{2 E^{\rm SM}}{\tilde{\lambda}_h },
\end{split}
\end{equation}
showing identical behavior, mainly controlled by the parameter $\tilde{\lambda}_h$ as in scenario A  above. 
Observe that the difference between the  $v_c / T_c $ expression in scenarios A and B, ~Eq.~(\ref{eq:a}) and (\ref{eq:1step}), is correlated with the difference between $\Delta V_{\rm A}$ defined in~Eq.~(\ref{eq:tildelh}), and the corresponding quantity for scenario B,
\begin{equation}
\begin{split}
\Delta V_{\rm B}  &\equiv V(0,0, T=0) - V(v_{\rm EW}, w_{\rm EW}, T=0) = \frac{v^4}{4} \tilde{\lambda}_h + \frac{(\mu_s^2)^2}{4 \lambda_s}.
\end{split}
\label{eq:tildelhB}
\end{equation}
Eq.~(\ref{eq:1stepaprox}) is a reflection that $\Delta V_{\rm A}$ and $\Delta V_{\rm B}$ only differ by the term $\frac{(\mu_s^2)^2}{4 \lambda_s}$, which again is small for Scenario B with SFOPhT.

The numerical results shown in Fig.~\ref{fig:pt}, highlight scattered points for scenario B (and scenario B-NR) in  green. According to  our discussion above, the quantity $v_c/T_c$ (upper panel) follows closely the expected behavior as a function of  $\tilde{\lambda}_h $, in a very good agreement with  Eq.~(\ref{eq:1stepaprox}). 
We observed that the data are scattered more downward compared with scenario A,  and this is due to the small correction from the additional second term in the denominator of Eq.~(\ref{eq:1step}).

\subsection{$Z_2$ Non-restoration scenarios}
\label{sec:nR}
 
In scenarios A and B discussed above, the phase transition, either one-step or two-steps, starts from the trivial phase $(0, 0)$ at high temperatures. Interestingly, it is also possible to consider  that 
the $Z_2$ symmetry is not restored at high temperatures.

Using the same high temperature approximation as in Eq.~(\ref{eq:vap}), 
when the coefficient $c_h$ is negative, $h$ will acquire a non-zero VEV at high temperatures, which has been recently discussed in \cite{Baldes:2018nel, Meade:2018saz}.  For $c_h$ to be negative, a relevant negative contribution to it from   $\lambda_m$ is required (see Eq.~(\ref{eq:htvcoe})), and this can be in general achieved in models with multiple singlets. However, since, in our case, we only have one singlet, such large negative contributions will require a large value of $\lambda_m$.
Thus, the electroweak symmetry is always restored at high temperatures $\left< h\right>^{\rm hT} = 0$ in our one-singlet extension of the SM.

With $\left< h \right> = 0$ at high temperatures, the singlet phase reads
\begin{equation} \label{hts}
\begin{split}
\tilde{w}(T) \equiv \left< s (T) \right>^{h = 0}  =  \left( \frac{-\mu_s^2 - c_s T^2 }{\lambda_s} \right)^{1/2}.
\end{split}
\end{equation}
For $c_s \ge 0$, with $\mu_s^2 \ge 0$, the phase $(0, \tilde{w})$ does not exist throughout the thermal history; while with $\mu_s^2 < 0$, the finite temperature  phase $(0, \tilde{w})$ can undergo $Z_2$ symmetry restoration  into the trivial phase $(0, 0)$ at a higher temperature 
 \begin{equation} \label{eq:tr}
\begin{split}
T^{Z_2}_r = \left( \frac{ - \mu_s^2}{c_s} \right )^{1/2}.
\end{split}
\end{equation}
The case $c_s \ge 0$ and $\mu_s^2 < 0$ is what drives Scenario A.  Observe that for Scenario B we also consider $c_s \ge 0$ (see Eq. (\ref{hts})), and because the transition is from $(0,0)$ to $(v,w)$, it also requires a positive defined  $T^{Z_2}_r$ 
 which in turn needs to be of  same order of the $T^{EW}_r \approx 140 \rm {GeV}$. Hence scenario B also requires $\mu_s^2 < 0$ as clearly shown in Fig.~\ref{fig:para}.
 
For $c_s < 0$, which can be achieved with negative $\lambda_m$, 
 Eq. (\ref{hts}) shows that 
the $Z_2$ symmetry remains non-restored at very high temperatures. This allows for thermal histories that start from a $(0, \tilde{w})$ phase and  can lead to extending scenarios A and  B to their $Z_2$ non-restoration corresponding cases. 
For both signs of  $\mu_s^2$, depending on its magnitude and the one of $c_s$,  one obtains the one-step phase transition that leads to scenario A-NR. If, however,  $\mu_s^2 \ge 0$, 
the $Z_2$ symmetry is temporarily restored at the temperature $T^{Z_2}_r$, given in Eq.~(\ref{eq:tr}), and it  is broken again to a different vacuum state,  $(v,w)$, during a later phase transition at a yet lower temperature. This is the path for scenario B-NR.

In summary, the novel condition of a SFOPhT with $Z_2$-NR explored in this work demands a  negative value of $c_s$, while different thermal histories are possible  depending on the value of $\mu_s^2$, as specify in Eq.~(\ref{cs_RvsNRconditions}) below and more clearly shown in Fig.~\ref{fig:para},
\begin{equation}
\begin{aligned}
 & \qquad {\rm Z}_2-{\rm R}:\ c_s \ge 0 \qquad \qquad\qquad\qquad\quad\quad {\rm Z}_2-{\rm NR}: c_s < 0. \\[6pt]
  &{\color{blue}{\rm A}}: (0, 0) \to (0, \tilde w) \to ( v , w) \quad \Longrightarrow \quad \quad {\color{blue}{\rm A}-{\rm NR}}: (0,\tilde w)\to(v,w) \\[6pt]
&   {\color{blue}{\rm B}}: (0, 0) \to ( v , w) \quad\qquad\quad\ \Longrightarrow \quad  {\color{blue}{\rm B-NR}}: (0,\tilde w)\to (0,0) \to(v,w).
 \label{cs_RvsNRconditions}
 \end{aligned}
\end{equation}

The correspondence between the restoration and non-restoration scenarios  is defined by  them sharing the same final path towards the electroweak physical vacuum.
%
Thus, the enhancement effects on the transition strength from the singlet contribution can be described in the same manner. This implies that the ratio of  $v_c/T_c$ for scenario A-NR is described by the same~Eq.~(\ref{eq:a}) as in the scenario A. Analogously,   $v_c/T_c$ for scenario B-NR is described by~Eq.~(\ref{eq:1step}),  that after simplification becomes~Eq.~(\ref{eq:1stepaprox}) as in  scenario B, and therefore the same result as for scenario A. This also agrees with
 the fact that $\Delta V_{\rm A} \approx  \Delta V_{\rm B}$ for the points with SFOPhT, as discussed before, and it is clearly apparent from~Fig.~\ref{fig:pt} where there is a significant  overlap of data points in the $v_c/T_c$ - $\tilde \lambda_h$ plane, both for  scenarios  A and B as well as for the $Z_2$ restoration and non-restoration cases.

\begin{figure}
  \centering
  \includegraphics[width =0.8\textwidth]{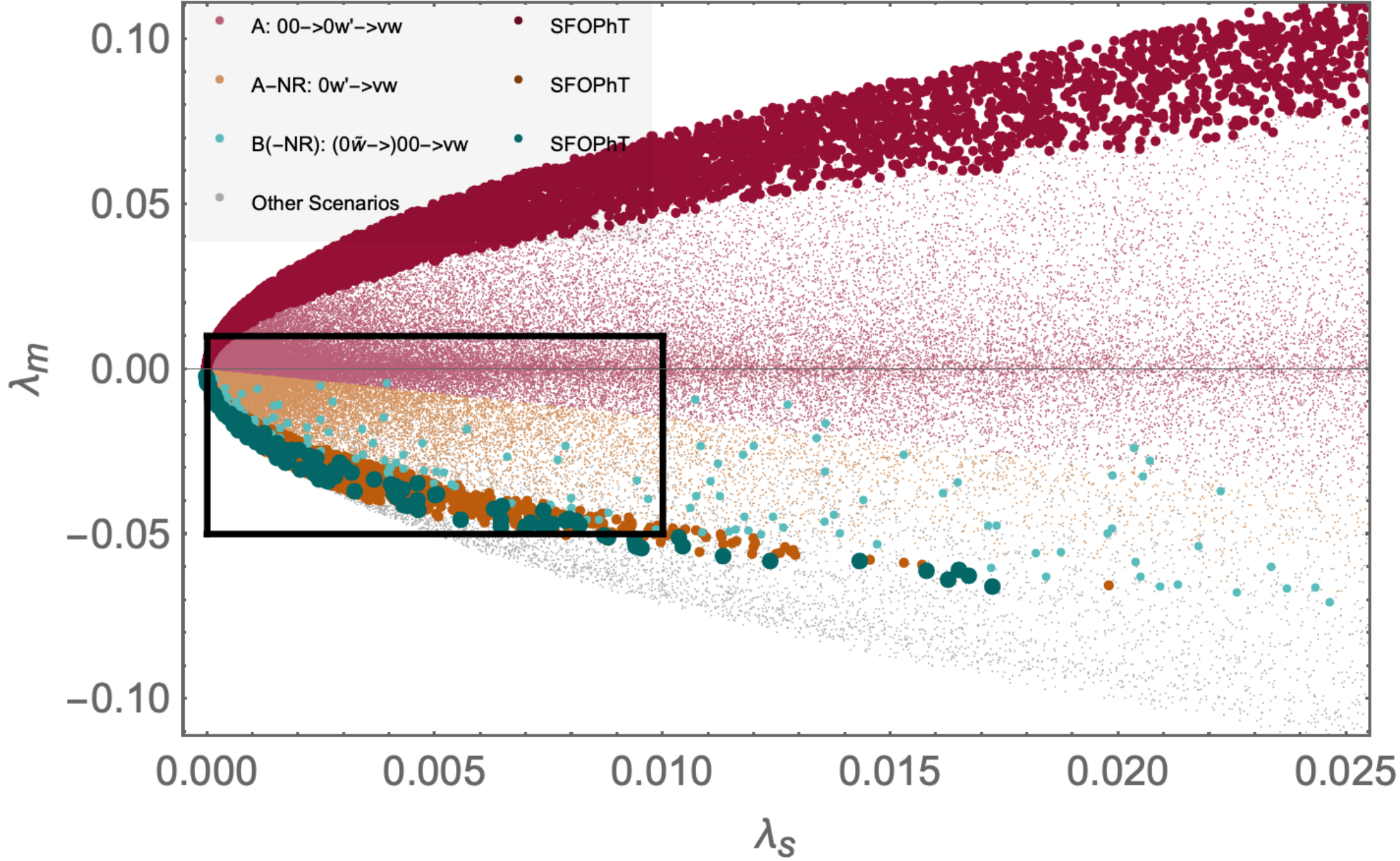}
  \caption{Parameter space on the $\lambda_s$-$\lambda_m$ plane with different phase transition scenarios, zoomed into the small $\lambda_s$ region. Color scheme for different scenarios is the same as in~Fig.~\ref{fig:para}.}
  \label{fig:lpara}
\end{figure}

The separation between the $Z_2$ restoration and non-restoration cases is clear in Fig.~\ref{fig:para}, corresponding to the positive and  negative $c_s$ regions, respectively. We have already described the restrictive region of scenario B. 
For scenario B-NR, $\sqrt{\frac{-\mu_s^2}{c_s} }$ is the temperature scale where $Z_2$ is temporarily restored from the high temperature $Z_2$ non-restoration phase, provided $\mu_s^2 > 0$. For a strong electroweak phase transition to happen in the step of $(0,0) \to (v, w)$ in scenario B-NR, this temperature needs to be below the $140$ GeV scale, i.e.~$\sqrt{\frac{-\mu_s^2}{c_s} }< 140$ GeV, otherwise after $Z_2$ symmetry restoration to the trivial phase, the transition to an electroweak breaking vacuum $(\tilde{v}, 0)$ will develop at a temperature around $140$ GeV, which will imply a small perturbation to the SM situation that we already know does not produce a SFOPhT. In addition, we expect this will result in scenario B-NR  transitioning  from $(0,0) \to (v,w)$  at a temperature significantly below 140 GeV, rendering a SFOPhT.
In  Fig.~\ref{fig:para}, this can be seen in the dark green shade points with negative $c_s$. Also observe from~Fig.~\ref{fig:para} that there is no SFOPhT points for the $Z_2$ restored scenario B.

In Fig.~\ref{fig:lpara}, we show the same data set as in Fig.~\ref{fig:pt} and Fig.~\ref{fig:para} for all the scenarios, now projected in the $\lambda_s$-$\lambda_m$ plane of the quartic couplings, zoomed into the small $\lambda_s$ region. As the Higgs quartic $\lambda_h$ varies within a small numerical range, the EWSB condition $\tilde{\lambda}_h \ge 0$ corresponds to the outer parabolic boundary of the dark region, and the SFOPhT condition $\tilde{\lambda}_h \lesssim 0.06$ corresponds to the inner parabolic boundary of the dark region. Different scenarios are coded by color in the same way as in  Fig.~\ref{fig:para} with dark shaded points corresponding to a SFOPhT. The points inside the rectangle are compatible with current bounds on the Higgs exotic decays, as will be discussed in Sec. \ref{sec:HiggsPheno}.
 
\subsection{Full one-Loop study and nucleation}
\label{sec:loop}

In this section, we shall show the results of the numerical scanning after implementing the CW and daisy resummation corrections introduced in Sec.~\ref{sec:thermal}.  All scanning results satisfy the Higgs mass and Higgs VEV boundary conditions. Other bounds will be introduced and shown in the following discussions. In Appendix~\ref{sec:scheme},  we will expand on details of the CW implementation, including discussions on the renormalization schemes, scale dependence, and Renormalization Group Equation (RGE) running.  

\begin{figure}
  \centering
  \includegraphics[width =0.5\textwidth]{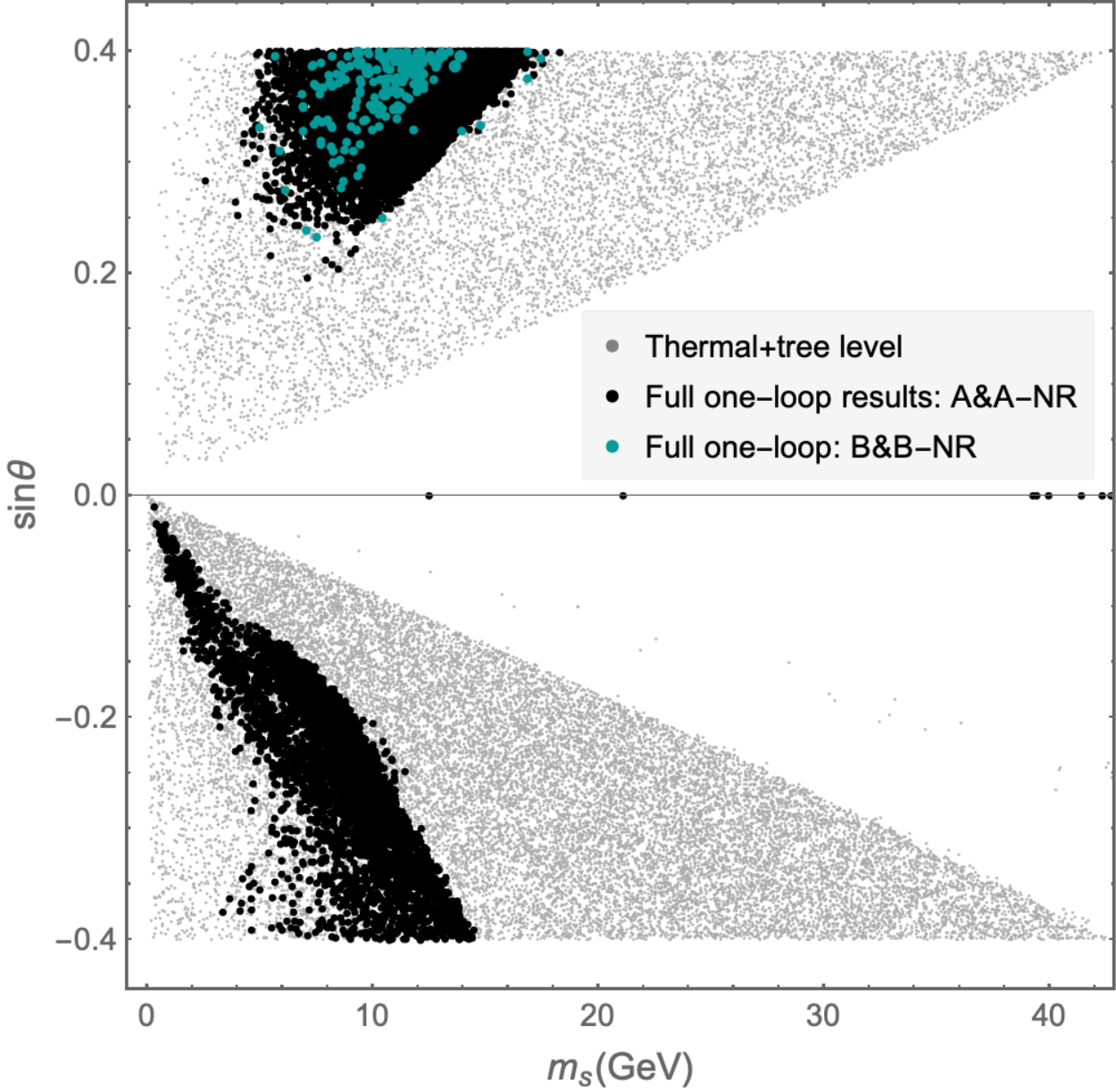}
  \caption{Parameter space for SFOPhT in the $m_S$ - $\sin\theta$ plane, after including  the full potential up to one loop order ( tree-level potential, one-loop thermal potential and one-loop zero temperature CW potential) plus finite temperature daisy resummation (darker shaded points in green and black for the B-NR and A/A-NR cases, respectively).  Also shown are the points with SFOPhT when only the  tree-level with one-loop thermal potential is considered (gray scattered points). 
  }
  \label{fig:1loop-para}
\end{figure}

Fig.~\ref{fig:1loop-para} shows the parameter space rendering SFOPhT after implementation of the full one-loop effective potential, including the one-loop thermal and CW potential, and the daisy resummation, projected on the physical parameter space of the singlet mass $m_S$ and the mixing angle $\sin\theta$. 
Observe that the sign of $\sin\theta$ is opposite of
the sign of $\lambda_m$ for values of $m_S < m_H$, as those of relevance in this study, see~Eq.~(\ref{eq:treeparaapp}). In addition, positive (negative) values of $\lambda_m$  are correlated to restoration (non-restoration) scenarios with SFOPhT (e.g.~see Fig.~\ref{fig:lpara}). As a result, it follows that all the solutions with SFOPhT and $\sin\theta < 0$ in Fig.~\ref{fig:1loop-para} correspond to the thermal history of scenario A (with $Z_2$ restoration), while solutions for  
$\sin\theta > 0$ in Fig.~\ref{fig:1loop-para} correspond to the thermal histories of scenarios A-NR (black) and B-NR (green), respectively. Our study shows that the valid parameter region rendering SFOPhT has been reduced after including the full one-loop results and features 
 smaller singlet mass values. 
 Importantly, including the full one-loop effective potential with daisy resummation still allows for all types of solutions that existed in the thermal only analysis.

The CW correction to the scalar potential effectively accounts for the one-loop running of the tree-level potential parameters
~\cite{Bando:1992np,Ford:1992mv,Carena:1995wu,Andreassen:2014eha,Andreassen:2014gha,Tamarit:2014dua}.  The top quark Yukawa coupling yields the most relevant contribution in the running of the quartic couplings, with the possibility of rendering them  negative  at large scales.
Furthermore, as we have discussed in detail in Sec.\ \ref{sec:SA}, the effective quartic $\tilde{\lambda}_h (v_{\rm EW})$, which is directly related to the phase transition strength, is required to be small to yield a SFOPhT. Hence the stronger the first-order phase transition, the smaller the effective quartic $\tilde{\lambda}_h (v_{\rm EW})$ and the most likely it is to be rendered negative at large scales, through the effects of the top Yukawa coupling in its running. This implies that after including the CW potential in the analysis, the points with stronger first-order phase  transition strength  in the thermal only analysis will be more likely to become unstable (acquire a negative effective quartic coupling) and will be discarded from the accepted solutions.  If instead, one would implement a RG improvement of the CW potential, this will include the effects of running of the top quark Yukawa coupling itself, diminishing its value at large scales and, hence, also its impact in rendering the effective quartic coupling unstable. As a result, the inclusion of the one-loop CW without the  RG improvement has the effect of reducing the parameter space of  SFOPhT as shown in~Fig.~\ref{fig:1loop-para}, beyond what would be the case with a more comprehensive analysis. In this sense the results presented in ~Fig.~\ref{fig:1loop-para} are conservative. We shall postpone a full study of the RG-improved effective one-loop scalar potential, as well as exploration of gauge dependence effects, for future work.

\begin{figure}
  \centering
  \includegraphics[width =0.5\textwidth]{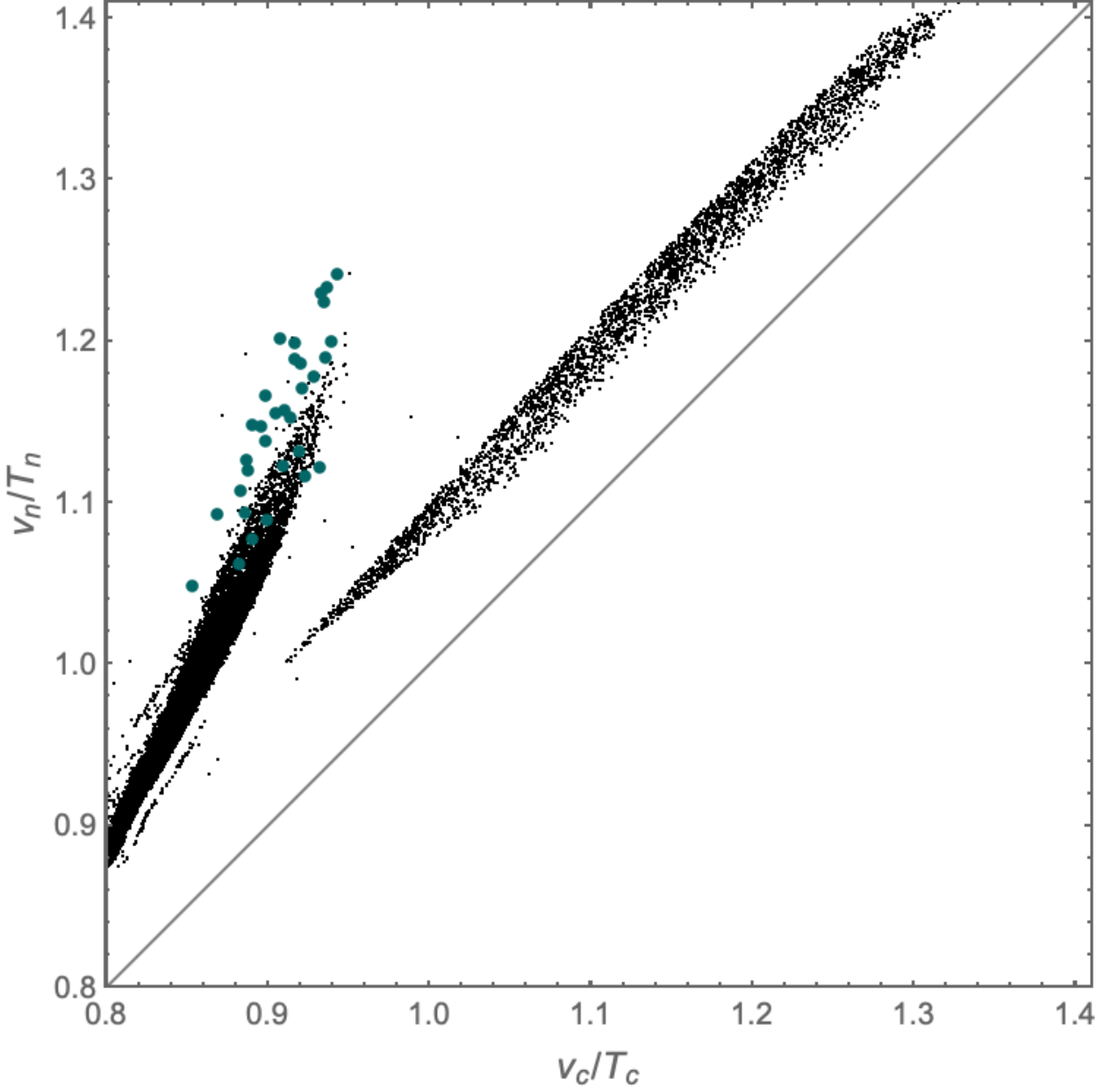}
  \caption{Nucleation calculation results at a full one loop level with daisy resummation corrections. Black: scenario A and A-NR. Green: scenario B and B-NR.
  }
  \label{fig:1loop_tn}
\end{figure}

It is crucial to check that our results are robust against the nucleation calculation.
Fig.~\ref{fig:1loop_tn} shows the nucleation calculation results including the full one-loop effective potential and the daisy resummation correction, we observe that the actual transition strength at nucleation temperatures is stronger than the strength evaluated at the critical temperatures. For computational efficiency, all the previous calculations have been done at  the critical temperature that gives a good  indication of the actual transition strength at the nucleation temperature. Therefore, Fig.~\ref{fig:1loop_tn} indicates that it is sufficient to require $v_c/T_c \gtrsim 0.8$ as criteria for a SFOPhT. 

\section{Phenomenology}
\label{sec:pheno}
The analysis of the thermal history of the spontaneous $Z_2$ breaking singlet extension of the SM leads to a firm prediction of a light singlet-like scalar mass eigenstate. The viable parameter space can be tested through various phenomenological probes. First of all, the spontaneous $Z_2$ breaking will result in mixing between the singlet scalar and the doublet Higgs boson. The Higgs precision measurements and electroweak precision measurements constrain the mixing angle $\sin\theta$ to be smaller than 0.4 for light singlets.\footnote{ The constrain improves to 0.2 for heavy singlets. For more details, see the appendix of Ref.~\cite{Carena:2018vpt}.} This constraint has been applied directly to our numerical scans. Furthermore, the precision Higgs program will improve with the full HL-LHC dataset~\cite{Cepeda:2019klc,CidVidal:2018eel}, and even more with data from future colliders~\cite{deBlas:2018mhx,Fujii:2017vwa,CEPCStudyGroup:2018ghi,An:2018dwb,Abada:2019zxq,Abada:2019ono,Benedikt:2018csr,deBlas:2019rxi}.

In this section, we discuss three leading observational aspects of the model in regions of parameter space compatible with a strongly first-order electroweak phase transitions (SFOEWPT). First of all, the 125~GeV Higgs-like boson can decay to a pair of singlet-like scalars that can be directly searched for at the HL-LHC and/or at a  future collider Higgs factory. Second, the Higgs trilinear coupling is modified when compared with the SM one. Third, the strongly first-order phase transition can be potentially probed by the next generation of gravitational wave detectors.
In the following discussions we do not attempt to disentangle  between the different possible thermal histories of the spontaneous $Z_2$ breaking singlet extension of the SM.
For the collider phenomenology one would need to identify the signal dependence on the sign of 
 the mixing angle $\sin\theta$. This would require to perform a more involved phenomenological study beyond the scope of this work. Such a study will be relevant in case high precision LHC data points towards a Higgs exotic decay signal and an anomalous Higgs trilinear coupling.

\subsection{Higgs exotic decays}
\label{sec:HiggsPheno}

\begin{figure}[htbp]
  \begin{center}
    \includegraphics[scale=0.7,clip]{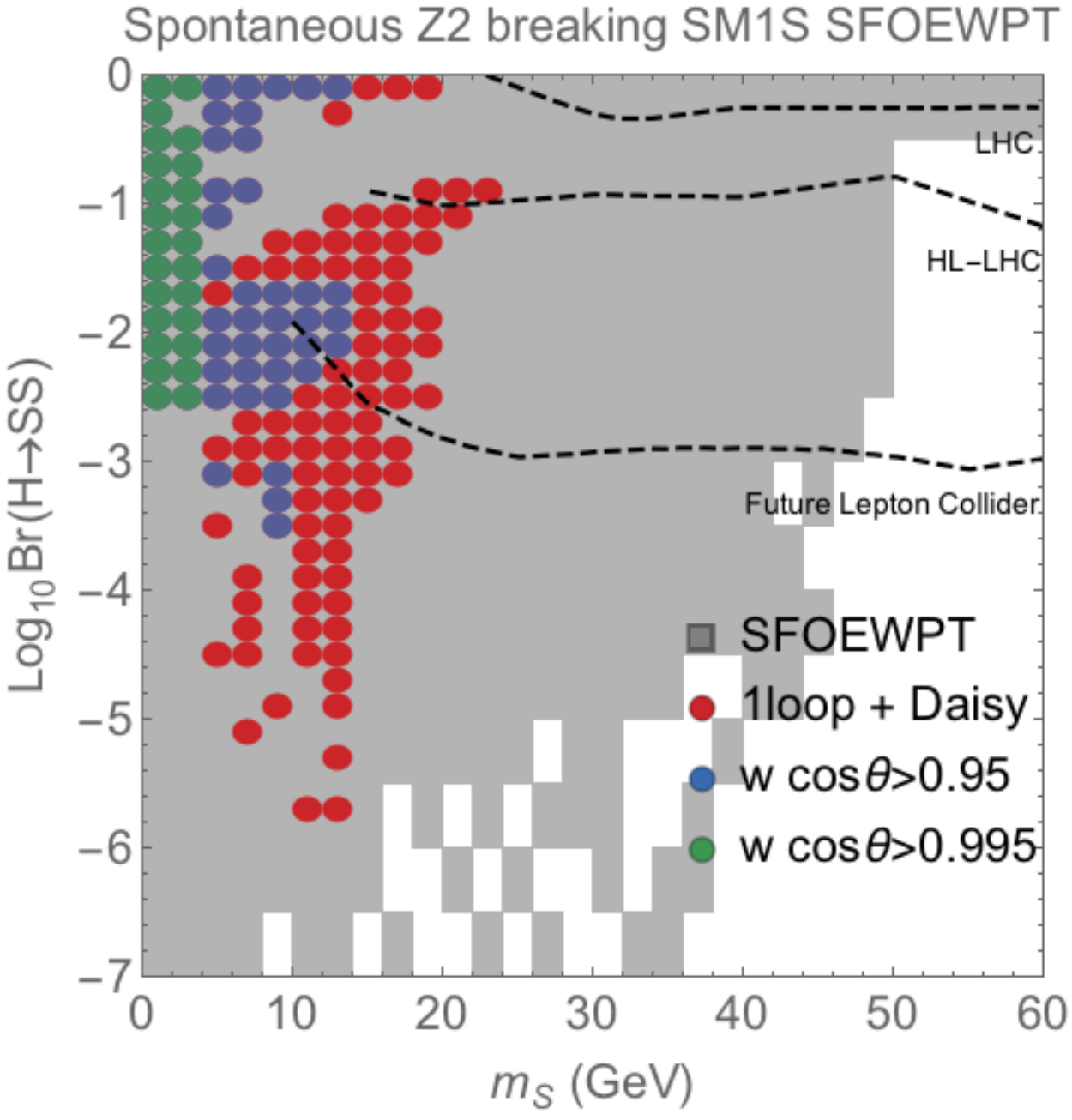}
    \caption{
    The Higgs decay branching fractions to $S$ pairs for points consistent with SFOEWPT, where $v_c/T_c \gtrsim 0.8$. The gray region includes one-loop thermal potential only. The red region in addition, include the one-loop CW potential and daisy resummation. The blue and green regions are compatible with $\cos\theta>0.95$, while the green region additionally requires $\cos\theta>0.995$, which are the HL-LHC and the future lepton-collider Higgs factory expected precision sensitivities on the Higgs-singlet mixing angle $\theta$~\cite{deBlas:2019rxi}. The upper and middle dashed lines define the lower value of the current and HL-LHC projected sensitivities to $H\to SS\to 4j$ searches. The lower dashed line corresponds to constraints from direct exotic Higgs decay searches at future lepton colliders~\cite{Liu:2016zki}.
    } 
  \label{fig:brss}
  \end{center}
\end{figure}

Since the singlet consistent with SFOEWPT should have a mass well below half of the SM-like Higgs boson one, the Higgs boson will decay into a pair of the new singlet scalars, $H\to SS$. The singlet-like scalar $S$ will then decay back to SM particles, dominantly into  a $b\bar b$ final state, if $m_S$ is greater than 10~GeV, and into other fermions and hadrons for lower singlet-like scalar masses~\cite{Curtin:2013fra}. The partial width of the SM Higgs decaying to the light singlet-like scalar $S$ is
\beq
\Gamma(H\to SS) = \frac {\Lambda_{HSS}^2} {32\pi m_H} \beta_S,
\eeq
where $\Lambda_{HSS}$ is the 
dimensionful coupling of the term $HSS$ in the mass basis. $\Lambda_{HSS}$ can be expressed as (without Coleman-Weinberg corrections),
\beq
\Lambda_{HSS}=\frac {(m_H^2+2 m_S^2)(- \cos\theta+\tan\beta \sin\theta)\sin2\theta} {4\tan\beta~v}.
\eeq
and $\beta_S=\sqrt{1-4m_S^2/m_H^2}$.

The current LHC Higgs exotic decay searches constrain the BR($H\to SS$)  to be smaller than around 25\% from a global fit~\cite{Khachatryan:2016vau,deFlorian:2016spz,Sirunyan:2018koj,ATLAS:2018doi} and 30-50\% from direct searches~\cite{Sirunyan:2018pzn,Aaboud:2018iil}. This translates into a constraint on  the $HSS$ coupling $\Lambda_{HSS}$ to be smaller than about 3~GeV. 
Given that for a large part of the parameter space, the size of this coupling reaches values up to $O(100)$~GeV, the Higgs exotic decay bounds provide an important constraint on this model. 

Fig.~\ref{fig:brss} shows the allowed values in the $\log_{10} BR(H \rightarrow SS) - m_S$ parameter space for different calculations of the SFOEWPT, with $v_c/T_c \gtrsim 0.8$.
The gray region includes only the tree-level and one-loop thermal contributions to the scalar potential.
The full one-loop results, including the CW corrections as well as the daisy resummation, are shown as the red, blue, and green regions for different requirements on the value of the Higgs-singlet mixing angle $\theta$. The HL-LHC Higgs precision measurements will be able to probe deviations of the Higgs boson couplings at the 5\% level, and this is shown by the blue and green regions. A future Higgs precision program at a prospective Higgs factory will measure the Higgs couplings at the  0.5\% level, which would limit the Higgs-singlet mixing angle $\cos\theta$ to be greater than 0.995, and is shown by the green region.\footnote{Note that the colored regions show allowed solutions without implying any assumptions on the density of such solutions, since this would be correlated to the density of scanned points, implying a highly prior dependent result. The same consideration is valid for~Fig.~\ref{fig:HHH}.}
Above the dashed lines in Fig.~\ref{fig:brss} are regions constrained by direct searches of the Higgs decaying to a singlet scalar pair: from top to bottom, the dashed lines represent the current LHC coverage, the corresponding HL-LHC coverage, and projections for a  future electron-positron collider~\cite{Liu:2016zki}, respectively. 
As shown in Fig.~\ref{fig:brss},
imposing the future Higgs precision bounds 
implies a strong preference towards low singlet masses, however, we expect that a more intense numerical scan targeted  to specific mass regions may expand the mass values allowed.\footnote{ Note that for these results on a five-dimensional parameter space, we performed scans with approximately $10^5$ CPU hours. We have a total of $10^7$ points, of which $10^5$ are compatible with SFOEWPT, and $10^4$ satisfy the current Higgs precision and exotic decay constraints.}
The boundary is also affected by the renormalization scale choice of the CW potential (see discussion in~Appendix~\ref{sec:scheme}). We argue that the HL-LHC will be able to actively probe a significant region of the SFOEWPT parameter space in a spontaneous $Z_2$ breaking singlet extension of the SM and that a future Higgs factory could compellingly test this model.

\subsection{Higgs pair production}
\label{sec:HiggsPheno}

The Higgs pair production process provides a unique handle in exploring the vacuum structure of the Higgs potential~\cite{Carena:2018vpt,Chen:2014ask,Dawson:2015haa}. The HL-LHC program can  probe the Higgs trilinear coupling  through  double Higgs boson production with an accuracy of  50\%~\cite{Cepeda:2019klc}, whereas it could be measured at the 
40\% level at a low energy lepton collider~\cite{DiVita:2017vrr}, and at the 5-7\% level at the FCC-hh~\cite{Benedikt:2018csr} as well as at CLIC~\cite{deBlas:2018mhx}.

\begin{figure}[htbp]
  \begin{center}
    \includegraphics[scale=0.6,clip]{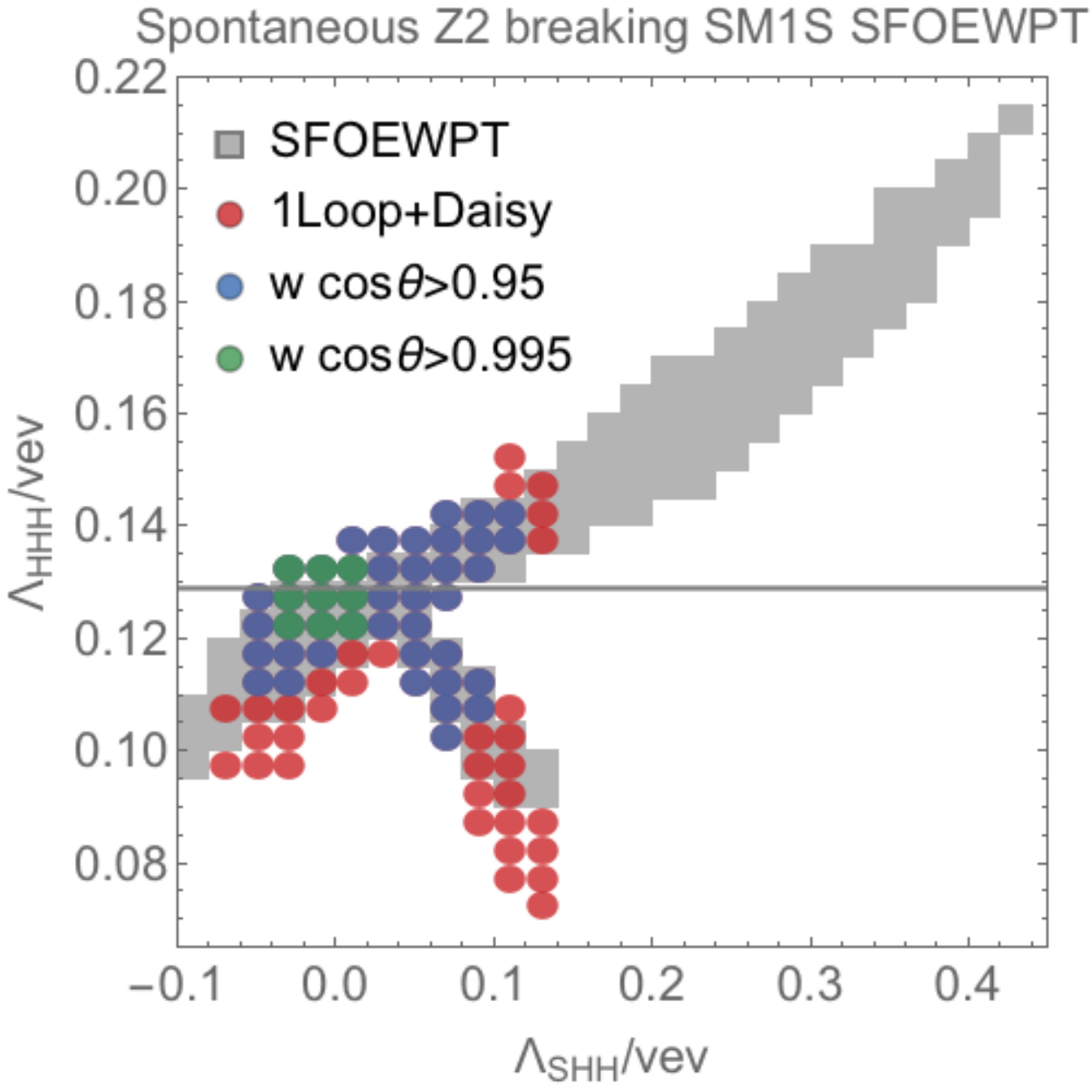}
    \includegraphics[scale=0.62,clip]{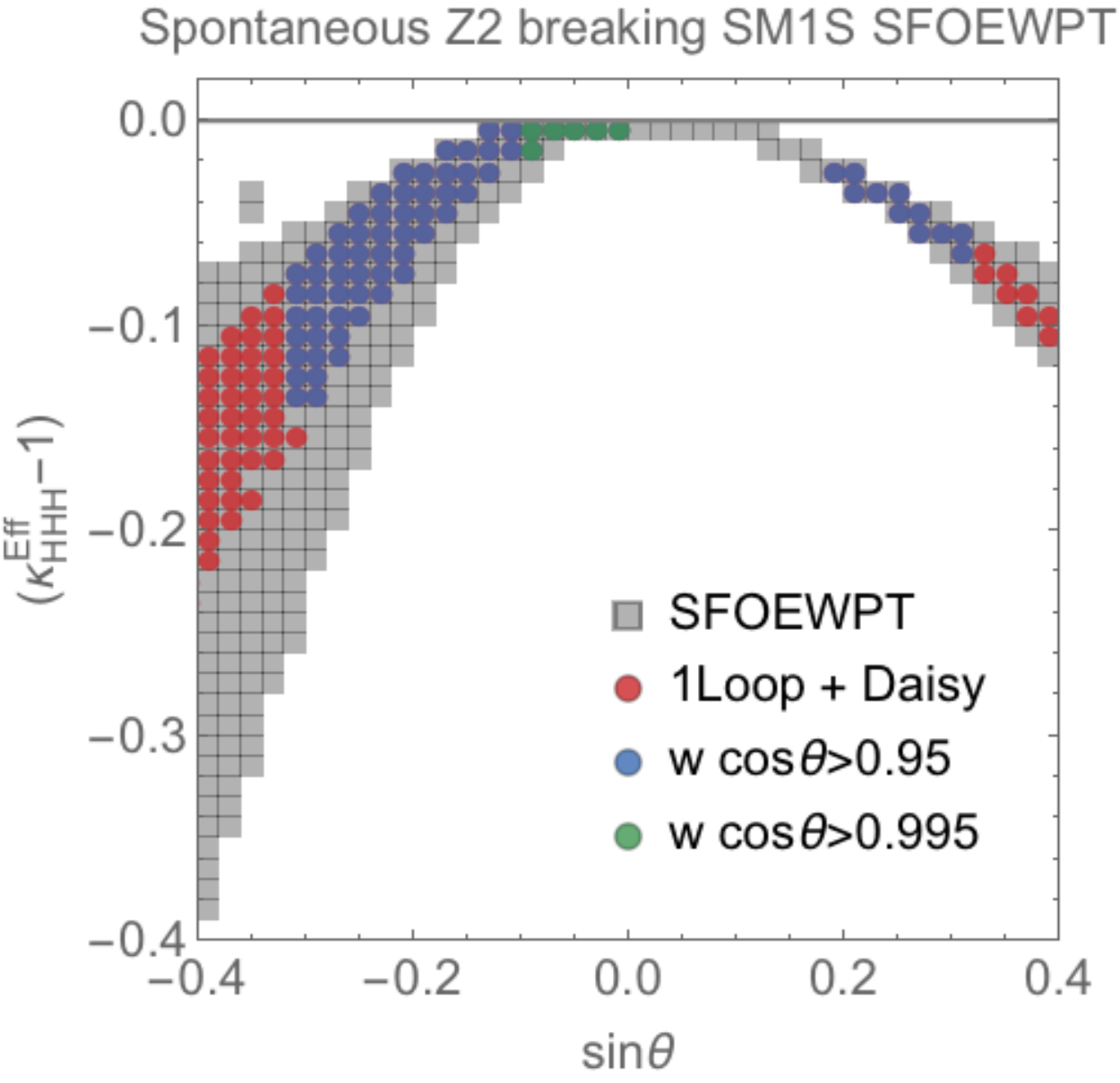}
    \caption{
   Left: Plane of the trilinear Higgs boson coupling and the singlet scalar-di-Higgs coupling, normalized to the SM Higgs boson VEV.    Right: Departure of the effective trilinear coupling $ \Lambda_{HHH}^{\rm Eff}$ from its SM value as a function of the mixing angle $\sin \theta$. For both figures the color coding is as follows:
   The gray region corresponds to results including the tree-level and one-loop thermal potential only. The red, blue and green disks, include the one-loop CW potential and daisy resummation. The blue and green disks further require $\cos\theta>0.95$, while the green disks additionally require $\cos\theta > 0.995$.
   The horizontal gray line indicates the SM value of  the y-axis parameter.
    } 
  \label{fig:HHH}
  \end{center}
\end{figure}

The Higgs pair production receives three contributions: the triangle diagram of an s-channel off-shell singlet $S$ through a $SHH$ vertex,  the triangle diagram of an s-channel off-shell $H$  through a $HHH$ vertex, and a top-quark box diagram  with double top Yukawa insertions. The first contribution from the s-channel off-shell scalar $S$ is additional to  the other SM ones, while the SM diagrams in turn are modified by mixing effects. The couplings governing the Higgs pair production are
\bea
\Lambda_{HHH}&=& \frac{m_H^2 \left( - \sin^3 \theta + \tan\beta\cos^3 \theta \right) }{2 \tan\beta~v} \nonumber\\
\Lambda_{SHH}&=& \frac {(2 m_H^2+m_S^2)(\sin\theta+ \tan\beta \cos\theta)\sin2\theta} {4\tan\beta~v} .
\label{eq:trilinear}
\eea

A further simplification can be made due to the fact that $m_S$ is much smaller than twice the Higgs mass. For the double Higgs production at hadron colliders such as the LHC and FCC-hh, within a  good approximation, one can define an effective trilinear coupling that combines the two triangle diagrams via
\bea
\Lambda_{HHH}^{\rm Eff}&=&\frac 2 3 \sin\theta \frac {\hat s^2} {(\hat s-m_S^2)^2+i \Gamma_S m_S} \Lambda_{SHH}+\cos\theta \frac {\hat s^2} {(\hat s-m_H^2)^2+i \Gamma_H m_H} \Lambda_{HHH}\\
&\simeq&\frac 2 3 \sin\theta \Lambda_{SHH} + \cos\theta\Lambda_{HHH}.
\label{eq:efftrilinear}
\eea
The determination and measurement of the trilinear Higgs coupling uses the differential information of the process as a result of the different diagrams and the interferences between the  SM di-Higgs box diagram and the effective triangle diagram. Indeed,  given the smallness of the singlet mass, the double Higgs production is far off-shell and can be absorbed into the above effective Higgs trilinear redefinition, which is valid at the differential cross-section level. 

We show the contributing trilinear couplings, $\Lambda_{HHH}$ and $\Lambda_{SHH}$, in the mass basis in the left panel of Fig.~\ref{fig:HHH}. The modified Higgs trilinear coupling $\Lambda_{HHH}$ varies broadly between 0.08 to 0.20. There is, in general, a positive correlation between $\Lambda_{HHH}$ and the singlet scalar-di-Higgs trilinear coupling $\Lambda_{SHH}$. Such a positive correlation follows from Eq.~(\ref{eq:trilinear}) for a subdominant contribution of the negative sin$^3\theta$ term in $\Lambda_{HHH}$, which corresponds to the mixing quartic coupling contribution. 
The case of negative correlation, instead, follows from the dominance of the negative sin$^3\theta$ term over the positive second term in $\Lambda_{HHH}$. The shading and color choices are the same as in Fig.~\ref{fig:brss}. We can see that as we restrict the Higgs-singlet mixing parameter $\sin\theta$ to be smaller, the Higgs trilinear coupling is also reduced to be closer to the SM value (which is shown as a gray reference line). 
The right panel of Fig.~\ref{fig:HHH} shows the departure of the effective trilinear coupling $ \Lambda_{HHH}^{\rm Eff}$ from its SM value as a function of the mixing parameter $\sin\theta$. We have defined the ratio
\beq
\kappa_{HHH}^{Eff}\equiv \frac {\Lambda_{HHH}^{Eff}} {\Lambda_{HHH}^{SM}},\nonumber
\eeq
with $\Lambda_{HHH}^{Eff}$  defined in Eq.~(\ref{eq:efftrilinear}) and, again, the color code  is the same  as  in Fig.~\ref{fig:brss}. We observe that for negative values of the mixing parameter $\sin\theta$, the effective Higgs trilinear coupling can be suppressed as much as 30\%, while for positive values, the suppression is at most of the order 10\%. These changes in the Higgs trilinear coupling are beyond the current reach of colliders and set a  compelling challenge for the di-Higgs boson search program and related precision measurements at future colliders. 



\subsection{Gravitational wave signature}
\label{sec:GRwave}

Discussions on Gravitational Wave (GW) signatures associated with a SFOEWPT in singlet extensions of the SM have been carried out in recent studies, see, e.g., Refs.~\cite{Huang2015,Kozaczuk:2015owa,Huang:2016cjm,Angelescu:2018dkk,Cheng2018,Alves:2018jsw,Alanne:2019bsm,Dev:2019njv}. Here we study for the first time the potential for detectability of gravitational waves in a singlet extension of the SM with 
spontaneous $Z_2$ breaking.
We provide a rough estimate of the GW signatures of the various underlying thermal histories and evaluate the opportunities to observe them at current and future GW detection experiments.

\begin{figure}[htbp]
  \begin{center}
    \includegraphics[scale=0.7,clip]{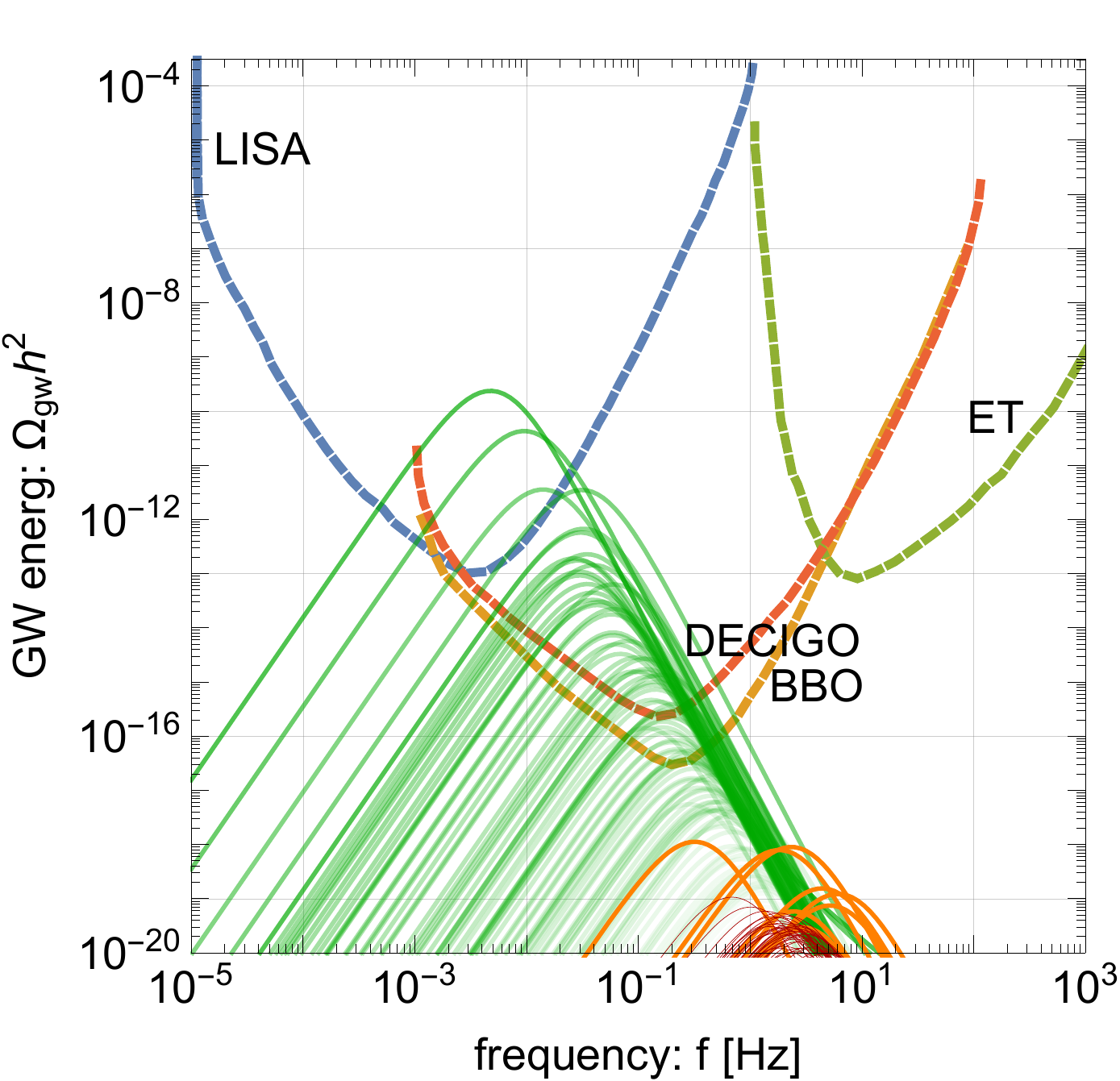}
    \caption{The gravitational wave power spectra generated during the  strong first-order EWPhT as a function of frequency for our model points. The green and orange curves correspond to  scenarios A and  B, respectively, including the tree-level and one-loop thermal calculation. The dark red curves correspond to the full 1-loop evaluation plus the daisy resummation. Also shown are the power-law integrated sensitivities of the LISA, BBO, DECIGO and ET projections, obtained from Ref~\cite{Breitbach:2018ddu}. The transparency of the green curves further indicates the strength of the EWPhT for the corresponding GW spectrum; the less transparent, the stronger the EWPhT.
    } 
  \label{fig:gw}
  \end{center}
\end{figure}

Following Ref.~\cite{Caprini:2015zlo}, we estimate the GW  signature spectrum from our model parameter points. The phase transition process induces the GW through bubble collision, dubbed as $\Omega_\phi$, propagation of the sound wave, dubbed $\Omega_{sw}$, and the decay of magnetohydrodynamic (MHD) turbulence, dubbed $\Omega_{MHD}$, respectively. The stochastic GW background power spectrum is the summation of these three sources,
\beq
h^2 \Omega_{GW}\simeq h^2 \Omega_\phi+ h^2 \Omega_{sw}+h^2 \Omega_{MHD},
\eeq
whose relative strengths differ depending on the given model. The detailed spectral and parametric dependences of the GW signature from the different sources are given in Ref.~\cite{Caprini:2015zlo,Caprini:2019egz}. In this section, we describe the key parameters and show the numerical results of our study.

The inverse duration of the phase transition is characterized by $\beta \simeq \dot{\Gamma}/\Gamma$ with $\Gamma$ the bubble nucleation rate. In turn, the relevant parameter for the GW signal is 
\beq
\frac {\beta} {H_{*}}\sim T \left.  \frac {d{(S_3/T})} {dT} \right|_{T=T_{*}},
\eeq
where $S_3/T$ is the $O(3)$-symmetric bounce Euclidean action, $T_*$ denotes the temperature of the thermal bath when the GW was generated, and $H_*$ is the corresponding Hubble parameter at temperature $T_*$. For a strongly first-order phase transition without significant reheating, $T_*$ is approximately the nucleation temperature $T_n$.
The nucleation temperature mainly spans over  the range of 50 to 100 GeV for all the scenarios considered in this study. However, for the thermal-only calculation of scenario A and A-NR, the nucleation temperature can extend to values as low as a few GeV.
The bubble collision generated power spectrum is suppressed by two powers of the duration of the phase transition, $H_*/\beta$, while the corresponding spectra generated by the sound waves and turbulences last longer and are suppressed by only one power of the duration of the phase transition. 
Another crucial characteristic parameter is the fraction of vacuum energy released during the transition
with respect to the radiation bath. Specifically, 
\beq
\alpha= \frac {\rho_{\tilde v, \tilde w}-\rho_{v, w}} {\rho_{rad}} |_{T=T_{*}},
\label{eq:alpha}
\eeq
where $\rho_{\tilde v, \tilde w}$ and $\rho_{v,w}$ are the zero temperature vacuum energy densities before and after the phase transition, evaluating the VEVs at the phase transition temperature $T_{*}$  . The radiation energy density, $\rho_{rad}$, is  approximately  given by $g_* \pi^2 T_*^4/30$, where $g_*$ is the number of relativistic degrees of freedom in the plasma at $T_*$. Note that as mentioned in Ref.~\cite{Caprini:2015zlo,Caprini:2019egz},
$\alpha$ also approximately coincides with the latent heat of the PT in the limit of a strong PT and large supercooling.

Another important parameter is the bubble wall velocity. If the wall velocity is small, then the GW spectrum is suppressed and hence less detectable. 
Detailed understanding of bubble wall velocity is, however, difficult, although one generically expects the plasma and matter reflection effects to let the bubble reach a relativistic terminal velocity~\cite{Caprini:2019egz}. In this study, we assumed it to be 0.5 of the speed of light, corresponding to non-runaway bubbles in the plasma.  In such a case, the energy from the scalar field is negligible, and the sound wave contribution to the GW signal dominates. Moreover,  we conservatively assume that the contribution from the MHD turbulence represents about 5\% of the total sound wave energy.

In Fig.~\ref{fig:gw} we show the GW spectral density associated to the strong first-order electroweak phase transition for various scenarios in comparison to the corresponding LISA~\cite{Audley:2017drz}, DECIGO~\cite{Seto:2001qf}, BBO~\cite{Corbin:2005ny} and Einstein Telescope (ET)~\cite{Sathyaprakash:2012jk} projected power-law integrated sensitivities~\cite{Moore:2014lga,Breitbach:2018ddu}. Other gravitational wave observatories such as Taiji~\cite{Guo:2018npi} and TianQin~\cite{Luo:2015ght} have similar sensitivities to LISA but different frequency bands, and future ones can further extend the gravitational wave sensitivities~\cite{Kawamura:2011zz,Crowder:2005nr,Baker:2019pnp}. The green and orange curves correspond to scenarios A 
and  B, respectively,  in the tree-level plus one-loop thermal calculation, for a sufficiently strong first order phase transition and allowing for nucleation. 
The dark red curves represent the  calculation including Coleman-Weinberg potential and daisy resummation, containing both scenario A  
and B points, again requiring nucleation. Given the  smallness and indetectability of these sets of signals, we do not distinguish among the different scenarios in the figure.
Observe that in our study, the temperature  of the thermal bath at the time when GWs are produced  is close to the nucleation temperature, since the  system  does not undergo a large supercooling, and hence we use $T_* = T_n$ in the calculations.
The GW signal is generated during the final or single step that gives place to the  EWPhT.  In the cases of a two-step phase transition (scenarios A and B-NR), the first step always involves a second order phase transition.

Now we comment on the several important aspects and underlying parameters of these GW spectra. For the green curves, the GW parameter $\alpha$ spans through the whole range from $5\times 10^{-4}$ to 100, \footnote{For strong-transitions,  with $\alpha$ of order 1 or larger, the dynamics of the GW becomes more complex and the GW strength computation has a large uncertainty that requires an improved treatment~\cite{Cutting:2019zws}. We note that the strongest four GW spectra shown by the green upper curves in Fig.~\ref{fig:gw} correspond to this case.} which strongly correlates with the strength of EWPhT. The larger the $\alpha$, the stronger the EWPhT and as well the lower the nucleation temperature. $\beta/H$ spans over  the range $5\times 10^2$ to $10^5$, with higher density of results around $10^3$. The $\beta/H$ anti-correlates with the strength of the EWPhT. We use the transparency of the green curves to indicate the corresponding strength of the EWPhT and to highlight the connection between the strength of the EWPhT and the parameters $\alpha$ and  $\beta/H$: the less transparent the green curves, the stronger the EWPhT.
 The stronger the phase transition, the lower the nucleation temperature and $\beta/H$, and the higher the value of 
  $\alpha$. These facts lead to the strongest GW spectrum with a peak frequency around 3~mHz. We observe that the nucleation calculation effectively removes many model parameter points in scenario B with strong EWPhT, and with our current statistics, scenario B spreads over $\alpha$ between 0.005 and 0.02, and $\beta/H$ between $7\times 10^4$ and $5\times 10^5$.  
For the full calculation with CW and daisy resummation, which corresponds to the dark red curves, $\alpha$ spans over the range $0.003$ to $0.007$ and $\beta/H$ over the range $3\times 10^{3}$ to $4\times 10^{5}$. The points are more scattered around due to the enhanced complexity of the model parameter space due to the higher order radiative corrections to the scalar potential, with no obvious correlations. As argued earlier, the CW plus daisy resummation calculation, as well as the nucleation calculation, leaves model points with higher nucleation temperature,  above 50~GeV, together with  higher value of $\beta/H$. This, in turn,  corresponds to a higher peak frequency at Hz level but suppressed strength of the GW signals for dark red and orange curves. 

Our present results show that  future GW experiments such as LISA and BBO would have limited  sensitivity to detect GW signals associated to the EWPhT in models with spontaneous $Z_2$ breaking.
As explained above, the dark red curves correspond to the full one-loop with daisy resummation study, and this renders a much weaker GW signal.
This is related to the loss of points with very strong first-order phase transition, that, in the tree level plus one-loop thermal analyses (green and orange curves),  are associated with smaller values of $\tilde \lambda_h$. In turn,   smaller values of $\tilde \lambda_h$ will be more likely to become unstable (acquire negative values with RG running  at large scales) and will be discarded from the accepted solutions.  We already discussed this in detail in\ Sec.\ \ref{sec:loop}.  The reason  that in Fig.~\ref{fig:gw} we are showing both sets, those with the tree level plus one-loop thermal  potential and those including  the full one-loop corrections with daisy resummation, is because we anticipate  that a further improvement in the computation of the scalar potential, namely the  RG-improved CW potential, will affect the running
 of $\tilde \lambda_h$  and  stabilize some of the discarded solutions, therefore  yielding  stronger gravitational wave signals.
  In fact, we expect that an RG improved effective potential treatment will yield results that lie somewhere in between the dark red and the thermal-only  contours. This will require an additional  comprehensive analysis that we postpone for  a future work.

 




\section{Summary and outlook}
\label{sec:summary}

The electroweak phase transition provides an appealing avenue for baryogenesis. In the SM, however, the electroweak phase transition turns to be a crossover instead of a first-order phase transition, as would be required to preserve a possibly generated baryon asymmetry in the presence of sufficient CP violation. In this paper, we study the simplest extension of the SM to render the EWPhT to be strongly first-order.
The SM singlet extension has been studied to great detail in the past, both in the $Z_2$ preserving scenario as well as for the case of a generic potential where $Z_2$ is explicitly broken. In our systematic study, we consider the unique scenario of spontaneous $Z_2$ breaking, including one-loop thermal effects with daisy resummation and the Coleman-Weinberg potential corrections. 
 We identify several very distinctive features of the spontaneous $Z_2$ breaking model:

 
\begin{itemize}

\item A variety of thermal histories can be generically achieved. We classify them according to the number of steps to achieve the EWPhT. We define scenario A (Sec.~\ref{sec:SA}) for $(0,0) \to (0,\tilde w)\to (v, w)$ (two steps) and scenario B (Sec.~\ref{sec:SB}) for $(0,0)\to (v, w)$ (one step). We also consider the possibility that at high temperatures there is non-restoration of the $Z_2$ symmetry and define scenario A-NR for $(0,\tilde w)\to (v, w)$ (one step), and scenario B-NR for $(0,\tilde w) \to (0,0)\to (v, w)$ (two steps)  (Sec.~\ref{sec:nR}). The relation between the restoration and non-restoration scenarios  is defined by them sharing the same final path towards the electroweak physical vacuum; 

\item Our study shows that scenario A, A-NR and B-NR lead to solutions with strong first-order EWPhT;

\item We derive simple analytical relations for such scenarios  and perform detailed numerical simulations. Our study 
has the potential to be  generalized to other scalar extensions of the SM with  novel  phenomenology, e.g., in the limit of EW symmetry non-restoration;

\item We find that in the spontaneous $Z_2$ breaking singlet extension of the SM, due to an upper bound on the singlet Higgs mixing quartic $\lambda_m$, the enhanced EWPhT can only be achieved via a particular scenario of nearly degenerate extrema. As shown in detail in this paper, having an extremum close in vacuum energy to the global minimum at zero temperature yields a low critical temperature and a large critical EW VEV, enabling strong first-order EWPhT. 
Furthermore, we check our results performing a nucleation calculation and found $v_n/T_n$ larger than $v_c/T_c$ for these solutions,  further validating our results;
\item The realization of a strong first-order EWPhT in this model predicts a light singlet-like scalar with a mass smaller than 50 GeV, which allows for  a rich phenomenology.  Special properties of the model can be tested through Higgs exotic decays and via  Higgs coupling precision measurements at current and future collider facilities. The trilinear Higgs boson coupling is modified and can be enhanced or suppressed with respect to its SM value. Future constraints on the Higgs boson self-coupling could shed light on the physics of the EWPhT. 
In addition, the strongly first-order EWPhT transition can generate gravitational-wave signals, which are a challenging target for future gravitational wave experiments such as LISA and BBO.

\end{itemize}

The above points summarize distinctive aspects of the spontaneous $Z_2$-breaking, singlet extension of the SM. The existence of a light scalar with accessible collider signatures as well as possible GW signals are common features that can also be present in more general models connecting the SM to a plausible dark sector via a Higgs portal.


\section*{Acknowledgment}
We thank J. Kozaczuk,  A. Long, M. Perelstein, M. Quiros, M. Ramsey-Musolf, J. Shelton and C. Wagner for helpful discussions.
We thank A. Long for his comments on our manuscript.
This manuscript has been authored by Fermi Research Alliance, LLC, under Contract No. DE-AC02-07CH11359 with the U.S. Department of Energy, Office of Science, Office of High Energy Physics. MC and ZL would
like to thank Aspen Center for Physics (Grant \#PHY-
1607611) and MIAPP program and ZL would like to thank IHEP, KITP, and KAIST programs for support and providing the environment for collaboration during various
stages of this work. ZL is supported in part by the NSF
under Grant No. PHY1620074 and by Maryland
Center for Fundamental Physics. 





\appendix
\section{Parameterization}
\label{sec:physpara}

There are five bare parameters in the tree-level potential, $\{\mu_h^2, \mu_s^2, \lambda_h, \lambda_s, \lambda_m\}$, that can be traded with five physical parameters $\{v_{\rm EW}, m_H, \tan\beta, m_S, \sin\theta\}$. The Higgs VEV $v_{\rm EW}$ and the Higgs mass $m_H$ are fixed by boundary conditions
\begin{equation} 
\begin{split}
v_{\rm EW}= 246\ {\rm GeV},\quad m_H=125\ {\rm GeV},
\end{split}
\end{equation}
whereas the remaining  three parameters are the  mass of the singlet-like eigenstate, the ratio of the singlet field VEV to the Higgs field VEV  and the mixing between the mass eigenstates, respectively:
\begin{equation} 
\begin{split}
m_S,\quad \tan\beta (\equiv \frac{w_{\rm EW}}{v_{\rm EW}}),\quad \sin\theta.
\end{split}
\end{equation}
The tree-level relations between these two sets of parameters are given by
 \begin{equation} 
\begin{split} \label{eq:treeparaapp}
&\mu_h^2=\frac{1}{4} \big(2 m_H^2 \cos^2\theta +2 m_S^2 \sin^2\theta + (m_S^2-m_H^2) \tan\beta \sin 2 \theta \big),\\
&\mu_s^2=-\frac{1}{4} \big(2 m_H^2 \sin^2\theta +2 m_S^2 \cos^2\theta + (m_S^2-m_H^2) \cot \beta \sin 2 \theta \big),\\
&\lambda_h =\frac{m_H^2 \cos^2\theta + m_S^2 \sin^2\theta}{2 v_{\rm EW}^2},\\
&\lambda_s =\frac{m_H^2 \sin^2\theta + m_S^2 \cos^2\theta}{2 \tan^2\beta v_{\rm EW}^2},\\
&\lambda_m =\frac{ (m_S^2 -m_H^2) \sin 2\theta}{2 \tan\beta v_{\rm EW}^2}.
\end{split}
\end{equation}
These tree-level relations provide a guidance for the understanding of the parametric dependence of the EWPhT strength, although such relations are modified after considering the CW corrections. 


\section{Aspects of the thermal potential}
\label{sec:VT}

The one-loop thermal potential reads
\begin{equation} \label{eq:VTA}
\begin{split}
V^T_{\rm 1-loop} (h, s, T)
&= \frac{T^4}{2\pi^2} \left[ \sum_{B} n_B J_B \left(\frac{m_B^2(h,s)} {T^2}\right) + \sum_{F} n_F J_F \left(\frac{m_F^2(h,s)} {T^2}\right)  \right],
\end{split}
\end{equation}
where $B$ includes all the bosonic degrees of freedom that couple directly to the Higgs boson, namely  $W, Z, \chi_i, \varphi_1, \varphi_2$ and $F$ stands for the fermionic particles that couple to the Higgs boson, but we are only considering the top quark. The $J_B$ and $J_F$ functions for bosons and fermions are defined as
\begin{equation} \label{}
\begin{split}
J_B (\alpha) = \int_0^{\infty} y^2 \ln{\Big[1-e^{-\sqrt{y^2 + \alpha}}\Big] dy},\\
J_F (\alpha) = \int_0^{\infty} y^2 \ln{\Big[1+e^{-\sqrt{y^2 + \alpha}}\Big] dy},
\end{split}
\end{equation}
and for positive values of $\alpha$ they have real values and are well defined. For negative $\alpha$ values, instead, the $J_B$ and $J_F$ functions become complex. As the effective squared masses in Eq.~(\ref{eq:VTA}) can be negative for some field values, we regulate the functions by taking their real parts \cite{Delaunay:2007wb}. After taking the real part, it occurs that for large values of $|\alpha|$, the $J$ functions have an oscillatory behavior around a central value. 
Numerically  {\tt CosmoTransitions} deals with such an oscillatory behavior by assigning  constant values to the functions once $|\alpha|$ becomes larger than  a certain large cut-off value, as it occurs when the system is  close to zero temperature.

All numerical studies in this work are based on results obtained with a modified version of {\tt CosmoTransitions} \cite{Wainwright:2011kj}, that appropriately accounts for our scenarios as well as  to improve on instabilities  under certain conditions. Our version of  the CosmoTransitions code is  available at \href{https://gitlab.com/yikwang/cosmotransition_z2sb.git}{CosmoTransitions-Z2SB}. For the numerical evaluation of the thermal potential with {\tt CosmoTransitions}, we chose  to  utilize the spline interpolation. 
We have compared the results obtained by performing the spline interpolation with those obtained by performing the exact integration of the $J$ functions. For all the benchmark points considered, we obtained  a reduction of at most $10\%$  in the strength of the phase transition. This difference is well within the limitations in accuracy due to the effective potential approximation we consider, namely one loop effective potential with daisy resummation.
Since the calculation with the spline interpolation is much more efficient than that one with  the exact integration, we used the former  throughout our study.

For the analytic evaluations  we  expand the  $J_B$ and $J_F$ functions  in terms of small $\alpha$, which yields the high-temperature approximation of the thermal potential. 
Under the high temperature expansion, the $J$ functions read 
\begin{equation} \label{}
\begin{split}
J_B^{\rm high-T} (\alpha) &={\rm Re} \Big[ -\frac{\pi^4}{45} + \frac{\pi^2}{12} \alpha - \frac{\pi}{6} \alpha^{\frac{3}{2}} + \cdots \Big],\\
J_F^{\rm high-T} (\alpha) &={\rm Re} \Big[ \frac{7 \pi^4}{360} - \frac{\pi^2}{24} \alpha + \cdots \Big].
\end{split}
\end{equation}
Based on the expansion (up to leading order in T), without the Coleman-Weinberg potential and daisy resummation contributions, the field-dependent part of the one-loop effective potential at finite temperature reads 
\begin{equation} \label{eq:htv}
\begin{split}
V(h,s, T) &= V_0 (h,s) + V_{\rm 1-loop}^{T} (h,s, T)  \\
&\approx - \frac{1}{2}(\mu_h^2 - c_hT^2) h^2  - E^{\rm SM}T h^3 + \frac{1}{4} \lambda_h h^4 \\
& \quad+ \frac{1}{2}( \mu_s^2 + c_s T^2) s^2 + \frac{1}{4} \lambda_s s^4 + \frac{1}{4} \lambda_m s^2h^2 - E(h,s) T,
\end{split}
\end{equation}
where 
\begin{equation} \label{eq:htvcoe}
\begin{split}
&  c_h \equiv \frac{1}{48} [9 g^2 + 3 g^{'2} + 2 (6 h_t^2 + 12 \lambda_h +\lambda_m)],  \\
&E^{\rm SM} \equiv  \frac{1}{32\pi} \Big[ 2 g^3 +  \sqrt{g^2+g^{'2}}^3 \Big] ,\\
&  c_s \equiv  \frac{1}{12} (2\lambda_m + 3 \lambda_s  ) ,  \\
& E(h,s) \equiv  \frac{1}{12\pi} \Big[   \big(m_{\varphi_1}^2(h,s)\big)^{3/2} +\big(m_{\varphi_2}^2(h,s) \big)^{3/2} +  3 \big(-\mu_h^2+ \lambda_h h^2 + \frac{1}{2} \lambda_m s^2 \big)^{3/2} \Big],
\end{split}
\end{equation}
where $m_{\varphi_{1,2}}^2$ is given in Eq.~(\ref{egm}).


\section{Aspects of the Coleman-Weinberg potential}
\label{sec:scheme}

As it is well known, it is possible to consider different schemes to evaluate the CW potential and each scheme has its own subtleties.  In particular, the  so called on-shell scheme is defined as the one that includes counter-terms such that the relations between the bare parameters in the Lagrangian and the physical parameters are not affected at one loop level.
This scheme has the advantage that in principle it has a fixed prescription for the renormalization scales for each particle and  that it is computationally more efficient, since the boundary conditions for the Higgs mass and VEV fixed at tree level remain valid at one loop level for the same set of bare parameters.
The CW potential in the on-shell scheme reads
\begin{equation} \label{}
\begin{split}
V^{\rm OS}_{\rm CW} (h,s)
= \frac{1}{64\pi^2} \sum_{i=\{B\},\{F\}} (-1)^{F_i}n_i \Big\{ & m_i^4(h,s) \Big[ \log \frac{m_i^2(h,s)}{m_i^2(v_{\rm EW},w_{\rm EW})}  - \frac{3}{2} \Big] \\
& + 2 m_i^2(h,s)m_i^2(v_{\rm EW},w_{\rm EW})  \Big\}.
\end{split}
\end{equation}
The on-shell scheme, however, has a subtlety related to the existence of massless particles at zero temperature. Indeed,  the Goldstone fields $\chi_i$ are massless at zero temperature and this yields  a CW potential in field space that is infrared divergent and ill-defined. Proper resummations should be employed to render the on-shell scheme consistent \cite{Elias-Miro:2014pca}.

In this study,  we have chosen to work in the $\overline{\rm MS}$ scheme, as has been introduced in~Sec.~\ref{sec:thermal}, to avoid the zero mass infrared divergency. However, in the  $\overline{\rm MS}$ scheme
the potential at the one loop level depends explicitly on the choice of the renormalization scale $Q$.
In our model, the singlet scalar could acquire a TeV scale VEV at zero temperature and hence we have chosen
 $Q = 1$ TeV throughout the study. 
To reduce the scale-dependence, an RGE improvement should be performed for the CW potential \cite{Bando:1992np,Andreassen:2014eha,Andreassen:2014gha,Tamarit:2014dua}. We leave the implementation of the RGE improvement  for future studies.

Note that, similar to the thermal potential, the CW potential is regulated by taking its real part in our study when the field-dependent squared masses become negative at some field values \cite{Delaunay:2007wb}.
 
\section{Other phase transition patterns}
\label{sec:exotics}


Another type of phase transitions that could occur  in the thermal history is
\begin{itemize}
  \item[]\begin{center}{(0,0)$\to$($\tilde v$,0)$\to$($v$,$w$)}\end{center}
    \item[]\begin{center}{(0,$\tilde w$)$\to$(0,0)$\to$($\tilde v$,0)$\to$($v$,$w$)},\end{center}
\end{itemize}
where the two scenarios differ from each other by the fact that  the $Z_2$ symmetry is restored or non-restored at high temperatures. Otherwise, they  share the same final path towards the electroweak physical vacuum. In both scenarios, the electroweak symmetry is first broken through the step 
\begin{equation}
\begin{split}
(0, 0) \to (\tilde{v}, 0),
\end{split}
\end{equation}
where $(\tilde{v}, 0)$ is an intermediate  phase at which the electroweak symmetry is broken while the  $Z_2$  symmetry remains preserved. Since the singlet does not acquire a VEV, it  plays no major role  in perturbing the potential depth at tree-level. Therefore the phase transition strength in this step is not largely affected by the existence of the extended singlet sector. Solving the finite temperature effective potential under the high temperature approximation, given in~Eq.~(\ref{eq:vap}), the strength of such a step is
\begin{equation}
\begin{split}
\frac{\tilde{v}(T_c) }{T_c}= \frac{2 E^{\rm SM}}{\lambda_h}  =  \frac{2 E^{\rm SM}}{\lambda_{h}^{\rm SM}} \Big[1 - \sin^2\theta \frac{m_S^2 -m_H^2}{m_H^2  \cos^2\theta + m_S^2 \sin^2\theta } \Big].
\end{split}
\end{equation}
The transition is enhanced when $m_S < m_H$. However, the enhancement is bounded from above by constraints from Higgs precision measurements, which roughly set the mixing angle $|\sin\theta| \lsim 0.4$. Accordingly, the transition strength is bounded as
\begin{equation}
\begin{split}
\frac{\tilde{v}(T_c) }{T_c} \le \frac{2 E^{\rm SM}}{\lambda_{h}^{\rm SM}} \Big[1 + \tan^2\theta  \Big] \lsim  1.2 \left( \frac{2 E^{\rm SM}}{\lambda_{h}^{\rm SM}}  \right) \approx 0.36.
\end{split}
\end{equation}
This upper bound on the transition strength is far below the requirement of SFOPhT. After including the CW potential and the daisy resummation corrections, such a step still yields small values of $\frac{\tilde{v}(T_c) }{T_c} $, provided the couplings still fulfill perturbative unitarity conditions.

\begin{figure}
  \centering
  \includegraphics[width =0.7\textwidth]{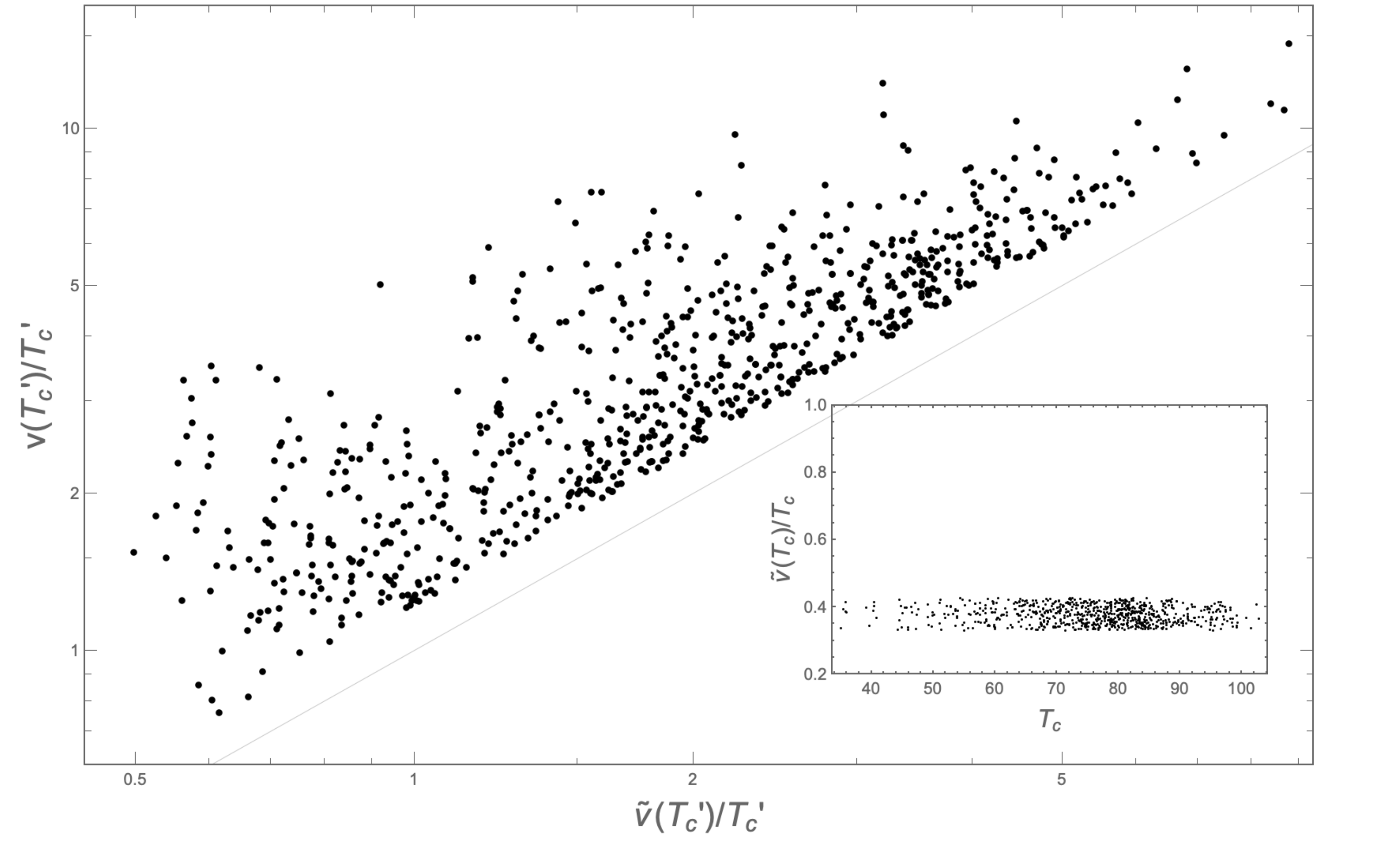}
  \caption{Higgs VEV to temperature ratios of the high temperature phase and low temperature phase for the transition step $(\tilde{v}, 0) \to (v, w)$ at a critical temperature $T'_c$. Results are obtained from numerical scanning with effective potential including tree-level and one-loop thermal potential. The sub figure shows phase transition strength of the previous step $(0,0) \to (\tilde{v}, 0)$ at a critical temperature $T_c$, where the electroweak symmetry is first broken and the Higgs obtains its non-zero VEV $\tilde{v}$. }
  \label{fig:00v0}
\end{figure}

From the temperature $ T_c$, at which the $(\tilde{v}, 0)$  is present, the thermal history proceeds to the next phase transition step, $(\tilde{v}, 0) \to (v, w)$, at a lower temperature $T'_c$, which breaks $Z_2$ and 
may further change the value of the  electroweak symmetry breaking vacuum. Such a step is either a smooth cross over, or a first-order phase transition. If it is of first-order nature,  the singlet field can play a role in rendering the strength of the phase transition strongly first-order. As shown in~Fig.~\ref{fig:00v0}, such is the case for $\frac{v(T'_c)}{T'_c} \gsim 1$ at $T'_c$, for which the sphaleron rate inside the bubble is suppressed.
However, we observe that the sphaleron rate outside the bubble is also suppressed during the bubble nucleation whenever the ratio of the high temperature phase $\frac{\tilde{v}(T'_c)}{T'_c} \gsim 1$. Therefore although the step $(\tilde{v}, 0) \to (v, w)$ can evolve a strongly first-order phase transition, no net baryon asymmetry can be created during the bubble nucleation.

In summary, although this type of thermal history occupies a sizable parameter space, it is not of special interests for modeling electroweak baryogenesis. The first electroweak breaking step $(0, 0) \to (\tilde{v}, 0)$ is weakly first-order, and any baryon asymmetry created is to be erased. For the following step $(\tilde{v}, 0) \to (v, w)$, 
although the phase transition can be strongly first-order, 
the sphaleron process is suppressed both inside and outside the bubble through the transition, therefore, no baryon asymmetry can be sourced.

\bibliographystyle{JHEP}

\bibliography{ewpht_sig}

\end{document}